\documentclass{article}

\usepackage{arxiv}

\usepackage[utf8]{inputenc} % allow utf-8 input
\usepackage[T1]{fontenc}    % use 8-bit T1 fonts
\usepackage{hyperref}       % hyperlinks
\usepackage{url}            % simple URL typesetting
\usepackage{booktabs}       % professional-quality tables
\usepackage{amsfonts}       % blackboard math symbols
\usepackage{nicefrac}       % compact symbols for 1/2, etc.
\usepackage{microtype}      % microtypography
\usepackage{lipsum}		% Can be removed after putting your text content
\usepackage{graphicx}
\usepackage{doi}
\usepackage{siunitx}
\usepackage[english]{babel}
\usepackage{amsmath,amssymb,amstext}
\usepackage{color}
\usepackage[backend=biber, style=nature, url=false,isbn=false]{biblatex}
\DeclareSourcemap{
	\maps[datatype=bibtex]{
		\map{
			\step[fieldset=abstract, null]
		}
	}
}

\title{Non-stationarity in correlation matrices for wind turbine SCADA-data}

\author{ Henrik M. Bette \\
	Fakultät für Physik\\
	University of Duisburg-Essen\\
	Duisburg, Germany \\
	\texttt{henrik.bette@uni-due.de} \\
	\and
	Edgar Jungblut \\
	Fakultät für Physik \\
	University of Duisburg-Essen\\
	Duisburg, Germany \\
	\texttt{edgar.jungblut@uni-due.de} \\
	\and
	Thomas Guhr \\
	Fakultät für Physik \\
	University of Duisburg-Essen\\
	Duisburg, Germany \\
	\texttt{thomas.guhr@uni-due.de} \\
}

\renewcommand{\shorttitle}{Non-stationarity in wind turbine correlations}

\hypersetup{
	pdftitle={Non-stationarity in correlation matrices for wind turbine SCADA-data},
	pdfsubject={},
	pdfauthor={Henrik M.~Bette, Edgar~Jungblut, Thomas~Guhr},
	pdfkeywords={},
}

\begin{document}

\maketitle

\begin{abstract}

Modern utility-scale wind turbines are equipped with a Supervisory Control And Data Acquisition (SCADA) system gathering vast amounts of operational data that can be used for analysis to improve operation and maintenance of turbines. We analyze high frequency SCADA-data from the Thanet offshore wind farm in the UK and evaluate Pearson correlation matrices for a variety of observables with a moving time window. This renders possible a quantitative assessment of non-stationarity in mutual dependencies of different types of data. We show that a clustering algorithm applied to the correlation matrices reveals distinct correlation structures for different states. Looking first at only one and then at multiple turbines, the main dependence of these states is shown to be on wind speed. This is in accordance with known turbine control systems, which change the behavior of the turbine depending on the available wind speed. We model the boundary wind speeds separating the states based on the clustering solution. Our analysis shows that for high frequency data the control mechanisms of a turbine lead to detectable non-stationarity in the correlation matrix. The presented methodology allows accounting for this with an automated pre-processing by sorting new data based on wind speed and comparing it to the respective operational state, thereby taking the non-stationarity into account for an analysis.
\end{abstract}

	% keywords can be removed

\section{Introduction}
Renewable energies are indispensable to respond to the temperature rise caused by climate change. Wind turbines are one key technology paving the way towards a green and emission-less energy production. The Global Wind Energy Council (GWEC) reports a growth of 93 GW in wind installations for 2020, bringing the total up to 650 GW with an increase of the growth rate by 53 \% compared to 2019. Offshore installations are on the rise, their comparatively low total increased from 29.1 GW by more than a fifth to 35.3 GW in 2019 \cite{Lee2021}. The GWEC also forecasts more than 205 GW new offshore wind capacity by 2030 in its 2020 Global Offshore Wind Report \cite{Lee2020}. They point out a growing acceptance of the fact that the price of offshore wind power can out-compete fossil and nuclear fuels. The various reports agree that wind power cannot be the only clean energy used to reach a net zero in emissions by 2050, but onshore and especially offshore wind turbines are crucial to reach this goal.

While offshore locations usually provide steadier and higher wind speeds, this comes at the cost of harsh environmental conditions and increased difficulties for operation and maintenance \cite{Feuchtwang2012}. Maldonado-Correa et. al. \cite{Maldonado-Correa2020} list various authors claiming that Operation and Maintenance (O\&M) costs account for 20-35\% of the total expenditure for offshore wind farms. The corresponding numbers are lower but still significant (approx. 10-15 \%) for onshore turbines \cite{Zhao2017, Tavner2012}.

Not surprisingly, among the many topics to be studied in the wind turbine field \cite{Veers2019}, O\&M is on focus for researchers and industry alike. They undertake increased efforts to effectively optimize O\&M procedures for wind turbines and thereby reduce this cost factor \cite{Novaes2018, Dao2021, Ulmer2020, Yan2019, Martin-del-Campo2020, Chen2021}. Improved understanding of wind turbine behavior is key to achieving this goal. Data driven methods are developed to control problems such as, for example, yaw misalignment or under-performance \cite{Pandit2018, Meyer2020}. Another prominent topic is the prediction of failures in wind turbines with sufficient lead time to react and carry out preemptive maintenance instead of correctional maintenance. This reduces not only the money lost in turbine downtime, but also enables cheaper maintenance. The idea is to optimize assets by replacing components exactly when needed \cite{Colone2018}. The wind energy branch follows a general trend in most industries, aiming at moving from scheduled maintenance towards condition-based maintenance to reduce costs and efforts \cite{Li2020}

A plethora of data are gathered in modern wind turbines. A Supervisory Control and Data Acquisition (SCADA) system is installed in all major wind farms since commission. This SCADA data contains many variables, usually averaged over 10 min intervals. Some further statistical measures, such as standard deviation in the 10 min interval, are often recorded as well. Many developed methods try to employ them for different types of analysis. The reader is referred to Maldonado-Correa et. al.\cite{Maldonado-Correa2020} and Tautz-Weinert et. al.\cite{Tautz-Weinert2017} for reviews. Common methodologies include neural networks, physical models and statistical analysis \cite{Jin2021, Kolumban2017, Gottschall2008, Aziz2021, Helbing2018, Pozo2018, Yan2019, Vidal2018, Marugan2019}. These authors also raise two important points: First, it is often complicated or impossible to reliably label events in the data due to scarcity of available log and maintenance data. Second, Ulmer et. al. \cite{Ulmer2020}, who apply convolutional neural networks for failure detection, mention that the 10 min averaging process naturally leads to a loss of information. This effect is specifically studied in \cite{Lichtenstein2021}. Some researchers have tried to avoid these problems by using simulated high frequency data \cite{Odgaard2013, Pozo2018} while the industry uses additionally installed vibration sensors to increase monitoring quality. Stetco et. al. \cite{Stetco2019} provide a review on approaches using such Condition Monitoring Systems (CMS). However, the goal is to reduce costs and installing additional sensors has its own inherent costs.

Another challenge for wind turbine analysis and monitoring is presented by the varying external (e.g. wind speed, temperature) and internal (e.g. turbine control, curtailment) operation conditions. Such non-stationarity is important also in many applications aside from wind turbines \cite{Kazemi2020, Shang2015, Canonaco2020, BuenoDeMesquita2021, Chen2019, Huang2022}. We have recently shown that non-stationarity in correlated systems is important for the detection of anomalies \cite{Bette2022, Tveten2019}. For wind turbines, it has been shown to have an effect on failure detection \cite{Zimroz2014, Bull2018}. Furthermore, different states in frequency data measured by a CMS system have been identified due to operational regimes \cite{Avendano2017}. Different behavior of the SCADA data for such regimes is to be expected also due to the turbine control mechanisms \cite{Novaes2018}. In general, complex systems containing many different variables, mechanisms and external influences show non-stationarity in their cross-correlations. The stability of correlations in financial stock market data was analyzed, for example, by Buccheri et. al. \cite{Buccheri2013}. Münnix et. al. \cite{Munnix2012} shows that the correlation matrix of this data inhabits different states over time by using cluster analysis. The stability of these states was further analyzed by Rinn et. al. \cite{Rinn2015} and Stepanov et. al. \cite{Stepanov2015}. Similar studies were also done for traffic data \cite{WangS2020}. Some general correlation analysis for wind turbine data was carried out by Braun et. al. \cite{Braun2020}.

In this paper, we aim to quantify the non-stationarity in correlation matrices for high frequency SCADA data from real wind turbines during normal operations. To this end, we apply cluster analysis to the correlation matrices of different SCADA signals calculated over 30 min time intervals. Distinct states with significantly different structures in the correlations matrices are found. We show that the prime cause for this is the turbine's own control mechanism. This allows us to develop a criterion based on wind speed to separate the cluster states. Such an automated distinction would, in principle, enable the usage of multiple normal states in applications via pre-processing. It is an important step towards accounting for the non-stationarity due to the operational regimes in an analysis such as failure detection with principal components.

This paper is structured as follows: In section \ref{sec:data} we will introduce the dataset we work with before moving on to the theoretical background of correlation matrices and clustering in section \ref{sec:theory}. Then we will present our clustering results for a single turbine in section \ref{sec:WT1Clustering}. We find proof of non-stationarity and identify the turbine control mechanism as the prime influence. Afterwards, we model the boundary wind speeds between the states in section \ref{sec:WT1Model}. Finally, we will show that the established method works for multiple turbines without further problems and can even be improved by the increase in available data in section \ref{sec:multiturbine}. We present our conclusions in section \ref{sec:conclusion}.

\section{Data}
\label{sec:data}
Our dataset includes 100 \emph{Vestas} V90 wind turbines from the Thanet offshore wind farm south-east of Great Britain. It contains observables that are measured at approximately 5 s time intervals. To obtain aligned, synchronized data, we average the time series on 10 s time intervals resulting in a data frequency $\nu = 1/10~\unit{s}$. This ensures continuity in the data even when the actual measurement frequency fluctuates around 5 s. This does not hinder the calculation of correlations for our purposes. In fact, it is rather similar to any measurement process: Every sensor will in reality average over a short time span to obtain a value. Taking the mean over 10 s seems therefore more natural than, for example, using only the last value from each interval. If at some time the deviation from the 5 s interval becomes stronger or if there are actually missing data points, data points will be missing in our averaged data as well. This might occur if there was a problem with, for example, sensors or communications. Another reason is simply a decreased measurement frequency when the turbine is switched off. This is underlined by the majority of missing values occurring in the very low wind speed regime beneath the turn-on wind speed. As the turbines are not running during these times, this wind range is not of interest for us. These missing values do not pose a problem for our analysis. For these reasons, we decided to transfer missing values from the original dataset into the 10 s data instead of replacing them by other means.

We are interested in identifying changes in the correlation structure, which emerge while the turbine is operating normally, in contrast to changes caused by failures. Therefore, in our main analysis we look at the following basic observables:
\begin{itemize}
	\item generated active power (ActivePower)
	\item generated current (CurrentL1)\footnote{As there are no deviations between the three phases in our data, we simply choose one of them.}
	\item rotation per minute of the rotor (RotorRPM)
	\item rotation per minute of the high speed shaft at the generator (GeneratorRPM)
	\item wind speed (WindSpeed)
\end{itemize}
These observables provide a good picture of the main turbine systems. Wind speed makes the rotor move. Its rotation is transmitted via gears to the rotation of the high speed shaft at the generator. This, in turn, generates electrical current and power. Two pairs of strongly correlated variables exist in this set. Deviations between generated active power and current could only occur, if large amounts of reactive power are generated. The low and high speed rotation of rotor and generator are directly coupled as well. These expected results are confirmed during our analysis. We include these pairs in our analysis as examples for group structures in the correlations. Knowledge of such structures is indispensable for monitoring a complex system: Are the groups stable? Do they break up? Do correlations across groups exist? In the presence of anomalies, such correlations, which are deemed normal and obvious, might be the structures which break up. Grouping is, in general, an important aspect in the study of complex systems \cite{Newman2012, Heckens2020a, WangS2020, Patankar2020, Toju2016}.

While measurements of temperatures are very common and useful for failure analysis \cite{Jin2021, Tautz-Weinert2017, Maldonado-Correa2020}, they are rather slowly changing variables. Hence, in our data they are not measured at high frequency and without decimals, which makes the calculation of short-term correlations impossible. Furthermore, it seems reasonable to assume that their behavior in normal states is strongly coupled to mechanical variables, e.g. higher rotation speeds will lead to increased bearing temperatures. Of course, they would also be influenced by, for example, seasonality or cooling mechanisms. Thus, while excluded from the study at hand, further analysis of temperature correlations is nevertheless desirable for future work.

Two additional important control variables are the pitch angle of the rotor blades (BladePitchAngle\footnote{Our data does not contain three separate measurements for the blades, but only one. We assume this to be a mean of the three individual blades.}) and the ratio between the blade tip speed and the current wind speed (TipSpeedRatio). The first is excluded in our main analysis due to many missing values that hinder the calculation of correlation matrices. The second is not directly present in the data, but results from easy linear calculation
\begin{equation}
	\label{eq:tipspeedratio}
	\text{TipSpeedRatio} = \frac{2\pi \text{RotorRPM} \cdot \text{RotorRadius}}{\text{WindSpeed}}.
\end{equation} It is disregarded in the main analysis, because we study linear correlations and it is also linearly derived from two already present variables. As both omissions are prominent observables when studying wind turbines, we include additional results with consideration of the pitch angle of the blades and the tip speed ratio for the basic cluster analysis in section \ref{sec:WT1Clustering}. To do this we had to fill the missing values for the pitch angle. A possible explanation for the missing values is a data acquisition system that only writes values whenever a new measurement is different from an old measurement, i.e. if the value changed. We could not establish whether this is actually the case in our data, but it seems reasonable. Thus, for the additional results including the pitch angle observable we treated the data as if this assumption were true. This means we filled any missing values in the 10 s data with the last measured value before. Of course, thereby we also fill in any values that might actually be missing instead of being left out for data storage reasons. Furthermore, with the pitch angle being a rather slowly changing variable compared to the other observables, the filling sometimes leads to stable values over long time periods. This, of course, hinders once again the calculation of short-term correlations. Therefore, one must be cautious when considering these additional results and we did not consider pitch angle or tip speed ratio directly for the analysis following the basic clustering in the present work.

We used approximately three weeks of data from 5 March 2017 to 24 March 2017. The data from such a time span are still easy enough to handle while providing enough data points to obtain reasonable clustering results. In view of possible practical applications, three weeks is a short enough time span to make it easily usable. There would be no need to collect huge amounts of operational data beforehand. However, it is of course necessary that the data used for identifying different operational states covers all possible states. In practice it turns out later that this means, we need a wide range of wind speeds in our data. The actual time span was chosen, because for at least one wind turbine there are no manual or automatic alarms or services during this period (cp. section \ref{sec:WT1Clustering}). Two turbines have no recorded data for this time span, effectively reducing our data set to 98 turbines.

Due to confidentiality agreements we will never show absolute values of any observable. In fact, only wind speed is shown directly and is then presented in units of the nominal wind speed $\tilde{v}_{\mathrm{nom}}$ at which the turbine starts to produce its nominal power output according to the manufacturer. The tilde is introduced to mark it as the rated value provided by the manufacturer as we will later on try to infer this value also from the data.

Each of the measured signals $k$ for each turbine $l$ yields a time series of data points $S_k^{(l)} (t)$, $k=1,\dots,K$, $l=1,\dots,L$, $t=1,\dots,T_{\text{end}}$. Here, we assume that all times are given as unit free steps. In the case of our dataset we have $K=5$, $L=98$ and - assuming complete data - $T_{\text{end}} = 20\cdot24\cdot60\cdot6=172800$. The data is arranged into $L$ rectangular $K \times T_{\text{end}}$ data matrices
\begin{equation}
	\label{eq:datamatrix}
	S^{(l)} = \begin{bmatrix}
		S_1^{(l)}(1) & \dots & S_1^{(l)}(T_{\text{end}}) \\
		\vdots & & \vdots \\
		S_k^{(l)}(1) & \ddots & S_k^{(l)}(T_{\text{end}}) \\
		\vdots & & \vdots \\
		S_K^{(l)}(1) & \dots & S_K^{(l)}(T_{\text{end}})
	\end{bmatrix},
\end{equation}
where each row is the time series of signal $k$.

\section{Theoretical background}
\label{sec:theory}

Our analysis to distinguish different states is based on identifying differences in the correlation matrix of observables listed in section \ref{sec:data}. In section \ref{sec:cormat} we define the way in which we are calculating the correlation matrices.

To identify non-stationarity in the time series of these matrices we will use a distance measure and a clustering algorithm. These are introduced in section \ref{sec:clustering}.

\subsection{Correlation matrices}
\label{sec:cormat}

To identify changes over time in the correlation structure, the correlation matrices are calculated with a moving time window of 30 min. The time intervals do not overlap. We effectively create a time series of matrices. To this end, the signal time series $S_k^{(l)}(t)$ are divided into disjoint intervals of 30 min, i.e. the lengths of the intervals is $T=180$ in our dimensionless time variable. We refer to these intervals as epochs. Hence, we have $T_{end}/T = 960$ epochs. To avoid notational confusion, we introduce the new time variable $\tau$ labeling the epochs. We reserve the notation $S_k^{(l)}(t)$ for the original time series and write $S^{(l)}(\tau)$ for the $K \times T$ data matrix containing the different time series for turbine $l$ from $\tau$ to $\tau+T-1$. The length of 30 min represents a compromise. Longer time spans would provide more data points per correlation matrix and would thereby decrease noise. However, we want to distinguish different states in time. Considering external conditions, e.g. wind, changing on short time scales of several minutes to hours, we have to choose relatively short epochs to ensure resolution of the non-stationarity. Such compromises are common when dealing with correlation matrix time series \cite{Marti2021, Heckens2020a}.

The first step towards calculating correlation matrices is the normalization of $S_k^{(l)}(t)$ to zero mean and standard deviation one in each epoch. With the mean value
\begin{equation}
	\mu_k^{(l)}(\tau) = \frac{1}{T}\sum_{t=\tau}^{\tau+T-1} S_k^{(l)}(t)
\end{equation}
and the standard deviation
\begin{equation}
	\sigma_k^{(l)}(\tau) = \sqrt{\frac{1}{T} \sum_{t=\tau}^{\tau+T-1} (S_k^{(l)}(t)-\mu_k^{(l)}(\tau))^2} ~~,
\end{equation}
of the epoch $\tau$, the normalized time series for signal $k$ and turbine $l$ is given by
\begin{equation}
	M_k^{(l)}(t) = \frac{S_k^{(l)}(t)-\mu_k^{(l)}(\tau)}{\sigma_k^{(l)}(\tau)} ~~, ~~ k=1,\dots,K ~~, ~~ l=1,\dots,L ~~,~~ \tau \leq t < \tau + T ~~.
\end{equation}
The normalized $K \times T$ data matrix $M^{(l)}(\tau)$ for each epoch is then defined analogous to $S^{(l)}(\tau)$.

The Pearson correlation matrix for each turbine $l$ during the time $\tau \leq t < \tau + T - 1$ is then easily calculated as
\begin{equation}
	C^{(l)}(\tau) = \frac{1}{T} M^{(l)}(\tau) M^{(l) \dagger}(\tau) ~~,
\end{equation}
where $M^{(l) \dagger}(\tau)$ denotes the transpose of $M^{(l)}(\tau)$. In the matrix $C^{(l)}(\tau)$ each element $C_{ij}^{(l)}(\tau)$ is the Pearson correlation coefficient of the signals $i$ and $j$ for turbine $l$ during the epoch from $\tau$ to $\tau+T-1$. The diagonal values are one by definition.

While the dependency of variables in a wind turbine is not always linear, which is already seen in the well-studied power curve, the linear Pearson correlation yields important and good results for the structure of the mutual dependencies. We have repeated our analysis with Spearman's rank correlation, which also measures non-linear dependencies, but did not find substantial differences. Results for the case with five variables are shown for comparison in appendix \ref{sec:app:Spearman}.
\subsection{Clustering}
\label{sec:clustering}

We will now introduce the clustering, which allows us to sort the correlation matrices into groups (clusters) and check, whether different typical states do exist. If we can identify these, we will refer to them as \emph{operational states}. An integer will be assigned to each of them and the algorithm will label each matrix in the time series with one such integer. Instead of a time series of correlation matrices, we then have a new integer time series $n(\tau)$ with the range $n \in \lbrace 1, \dots,  N \rbrace$ when $N$ is the number of clusters created.

The first outcome will then be as follows: If any decent clustering solution can be found, it is proof that typical states of the correlation structure exist. Then, analyzing the resulting integer time series $n(\tau)$ can much easier reveal dependencies of the state on time or other factors.

Any method separating objects into groups needs a distance measure defined between those objects. For the correlation matrices we choose the euclidean distance \cite{Heckens2020a}. The reader can imagine that all matrix entries are written into a vector, effectively arranging the columns of the matrix underneath each other, so that the standard euclidean distance between vectors can be applied. The distance between the correlation matrices for the epochs starting at $\tau$ and $\tau'$ of turbine $l$ is then
\begin{equation}
	\label{eq:distancemeasure}
	d^{(l)}(\tau, \tau') = \sqrt{\sum_{i,j}(C^{(l)}_{ij}(\tau)-C^{(l)}_{ij}(\tau'))^2} = ||C^{(l)}(\tau) - C^{(l)}(\tau')|| ~.
\end{equation}
We choose the bisecting $k$-means algorithm to perform our clustering \cite{Steinbach, Tan2019}. This is the algorithm that was also used to determine states in the financial markets by \cite{Munnix2012} and \cite{Heckens2020a}. It can be described as a hybrid of standard $k$-means clustering \cite{Lloyd1982, Jain2010} and hierarchical clustering \cite{Kaufman1990}. While the former directly divides the whole set of objects into $k$ groups (see appendix \ref{sec:standardkmeans}), the latter is performed step-wise. In each step either two groups are merged (agglomerative) or one group is divided into two (divisive). Bisecting $k$-means is a divisive clustering algorithm, meaning that at the start all objects belong to one big cluster and during each step one of the existing clusters is split into two new clusters. This bisection is performed by running a standard $k$-means on all objects within the group to be split with $k=2$. Which of the $\tilde{N}$ currently existing clusters is split during a step, is decided based on the average internal distance of all objects in a cluster $z_n ~, n=1,\dots,\tilde{N}$
\begin{equation}
	d^{(l)}_n = \frac{1}{|z^{(l)}_n|} \sum_{\tau \in z^{(l)}_n} ||C^{(l)}(\tau) - \langle C^{(l)} \rangle_n || ~~.
\end{equation}
Here, $|z^{(l)}_n|$ denotes the number of objects in cluster $z^{(l)}_n$ and $\langle C^{(l)} \rangle_n$ is the centroid of cluster $z^{(l)}_n$ defined by the element-wise mean:
\begin{equation}
	\label{eq:centermatrix}
	\langle C^{(l)}_{ij} \rangle_n = \frac{1}{|z^{(l)}_n|}\sum_{\tau \in z^{(l)}_n} C^{(l)}_{ij} ~~.
\end{equation}
Each step the cluster with the largest average internal distance is bisected. The algorithm is terminated when a set number $N$ of clusters is reached.

This is slightly different from the approach used in former works by \cite{Munnix2012}, \cite{Rinn2015}, \cite{Stepanov2015} and \cite{Heckens2020a}, where a threshold is introduced and all clusters are bisected until no single existing cluster has an average internal distance larger than the threshold. However, the threshold is then set based on the number of clusters wanted. It can easily be understood that the resulting clustering will be the same if one either uses our approach to produce $N$ clusters or chooses the threshold in such a way that $N$ clusters are produced.

Applying this algorithm, we split the set $Z^{(l)}=\lbrace C^{(l)}(1), C^{(l)}(1+T), \dots, C^{(l)}(T_{\mathrm{end}}-T) \rbrace$ of all correlation matrices into $N$ subsets $\lbrace z_1^{(l)}, \dots, z_n^{(l)}, \dots, z_N^{(l)} \rbrace$. The centroid of each cluster according to eq. \eqref{eq:centermatrix} is interpreted as the mean correlation matrix of a cluster representing its typical correlation structure. Thereby, we only need to look at $N$ matrices and a series of $T_{\mathrm{end}}/T$ integers $n(\tau)$ instead of as many matrices.

Later on we will see that the emerging typical correlation matrices correspond to different control settings of the turbines. This explains in a simplified way, why the hierarchical $k$-means works better than a normal $k$-means in our case. Approximately, we can describe the controller of a wind turbine as a mechanism fixing certain signals to a fixed value. This means the correlation of that signal with other non-fixed signals should vanish. The divisive clustering will first extract a group where signal $A$ might be fixed. Then this group might be further divided into subgroups where signal $B$ is either fixed or not. And in theory this could go on. Such a problem is very well suited for divisive clustering.

In order to check, if our clustering is sensible, we will do two things. Firstly, we will just look at the cluster centroids and see, if we can interpret them and if they are substantially different from each other. Secondly, we will calculate silhouette coefficients \cite{Rousseeuw1987}
\begin{equation}
	s^{(l)}(\tau) = \left\{
	\begin{array}{ll}
		\frac{b(\tau)-a(\tau)}{\max(b(\tau), a(\tau))} & , ~|z_{n(\tau)}| > 1 \\
		0 & , ~|z_{n(\tau)}| = 1
	\end{array}
	\right. 
\end{equation}
with the average distance to all other matrices in the same cluster
\begin{equation}
	a(\tau) = \frac{1}{|z_{n(\tau)}|-1} \sum_{\tau' \in z_{n(\tau)}, \tau' \neq \tau} d^{(l)}(\tau, \tau')
\end{equation}
and the smallest average distance to a single other cluster
\begin{equation}
	b(\tau) = \min_{m \neq n(\tau)}\left( \frac{1}{|z_m|} \sum_{\tau' \in z_m} d^{(l)}(\tau, \tau') \right) ~.
\end{equation}
This coefficient will take values between $-1$ and $+1$ with larger positive values representing matrices that are well clustered and negative values showing matrices that are closer to another cluster than to their own. To get an indicator for the overall clustering we will use the average silhouette coefficient
\begin{equation}
	\label{eq:avgsil}
	S^{(l)} = \frac{T}{T_{\mathrm{end}}} \sum_{\tau} s^{(l)}(\tau) ~.
\end{equation}

\section{State identification via clustering}
\label{sec:WT1Clustering}

In the following, we analyze the correlation matrix time series of one turbine, which will henceforth be referred to as turbine 1 (WT1). We are singling out this wind turbine, because the time frame of the total dataset was selected in such a way that for WT1 no problems were listed in the automatic alarm logs and manual service reports. The idea is that this will make analysis and definition of normal states easier as there was no (reported) unusual behavior.

The correlation matrices are calculated for non-overlapping epochs of 30 minutes each. This results in 960 matrices per turbine. However, due to several reasons there might be missing data in the time series. In such a case, any time stamp missing one or more of the measured variables is excluded from the calculation of the correlation matrix. Hereby, no estimation of values, which could influence the actual correlation coefficient, is necessary. More data could be used by calculating the correlation coefficients pair-wise, i.e. for any two observables just remove the time stamps where one of them is missing data. However, this does not result in a well defined positive semi-definite correlation matrix. For further analysis only those matrices, for which at least half of the expected data points (90 out of 180) exist, are considered. Furthermore, epochs in which the standard deviation $\sigma_k(\tau)=0$ for any signal $k$ have to be disregarded as they cannot be normalized.

We will first provide extensive results for the five considered observables in section \ref{sec:WT1ClusteringMain} and afterwards repeat some analysis including the pitch angle and tip speed ratio in section \ref{sec:WT1ClusteringAdditional}.

\subsection{Main results for five observables}
\label{sec:WT1ClusteringMain}
Without pitch angle no epoch includes a time series $k$, for which the standard deviation becomes zero. The disregarding of epochs with too many missing values is not a problem when looking at the five main observables as missing measurements usually stem from turbines being operational but switched off during times of very low wind speeds $v$ smaller than turn-on wind speed $v_{\mathrm{on}}$. For WT1 the average wind speed for 746 epochs with enough data is 10.01 $\unit{ms^{-1}}$, while the average for 214 epochs where no correlation matrix could be calculated is only 4.34  $\unit{ms^{-1}}$. It is obvious that these times where a wind turbine is not operating are unsuited for an identification of operational states. Of course, there might also be other reasons causing the missing data, e.g. a problem with the measurement of a signal. However, as for WT1 there are no alarms or services logged, we would not know what happened in those cases anyway. Any estimation of missing values would therefore need considerable guessing. As our results show that using only the epochs with enough data points is sufficient to reach a good differentiation of operational states, we are confident that just excluding missing values instead of estimating them is a good approach for our purposes.

\begin{figure}
	\centering
	\includegraphics[width=0.80\textwidth]{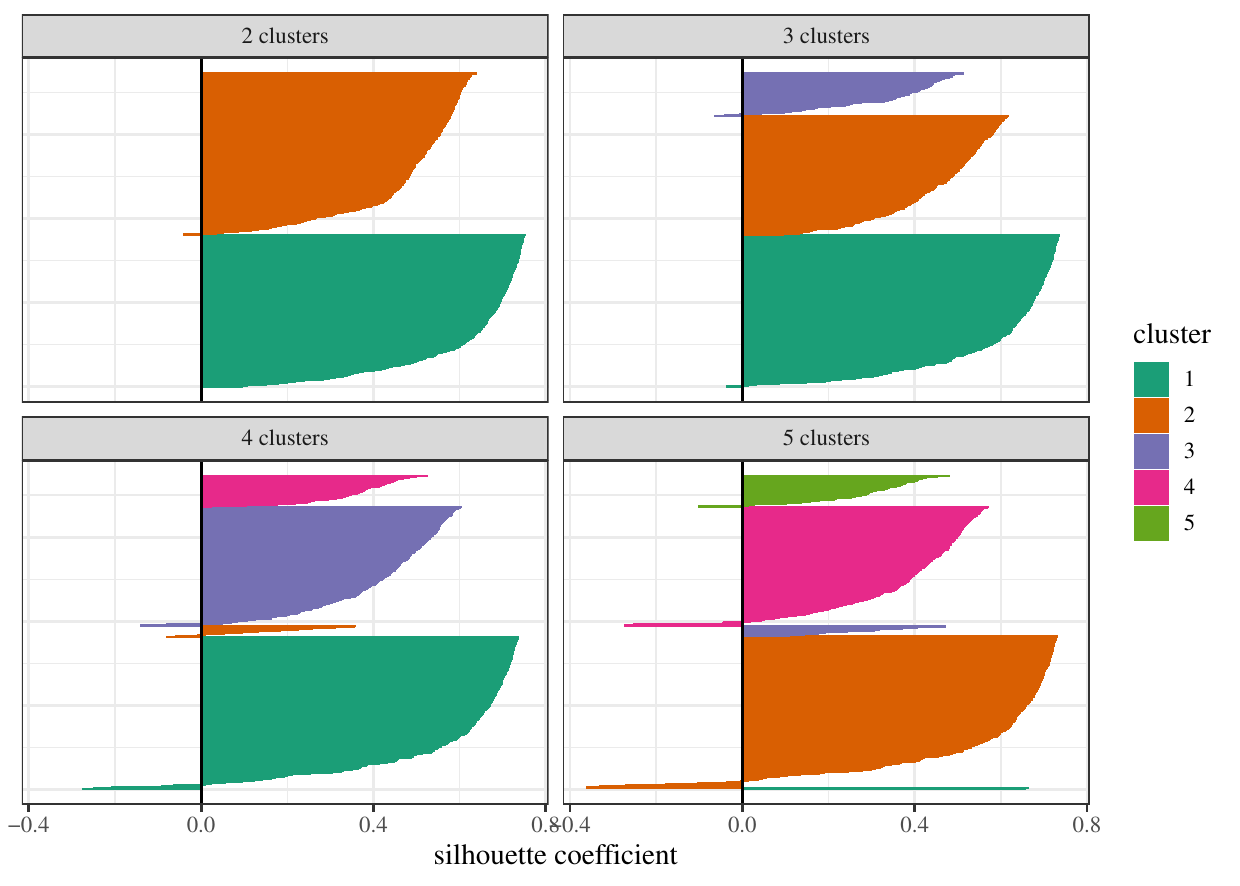}
	\caption{Silhouette plots for clustering solutions with 2-5 clusters. Each clustered element (matrix) is represented by a horizontal line the length of which is the silhouette coefficient for that element. Different clusters are color coded.}
	\label{fig:silhouettes}
\end{figure}

When applying the hierarchical $k$-means algorithm described in the previous section to the set of matrices, the first step is to decide how many clusters provide a good solution. To this end, we calculate the silhouette coefficients for solutions with 2-5 clusters. The silhouette plots can be found in fig. \ref{fig:silhouettes}. The fifth cluster is almost imperceptible as it consists only of 3 matrices. Some descriptive statistics for these silhouette coefficients are shown in table \ref{tab:silcoefs}. The \emph{mean} corresponds to the average silhouette coefficient from eq. \eqref{eq:avgsil}. All statistics provided decrease with increasing cluster number, implying that a few different states are sufficient to describe the dynamics of the analyzed correlation matrices. In the plots we can see some negative coefficients implying elements that would fit better into a different cluster. Such imperfections are to be expected when using heuristics like clustering algorithms. It is however clear, that all solutions provide a good grouping with largely positive silhouette coefficients. This is a strong indication that non-stationarities influence the correlation matrix. The influence is strong enough to be detected via simple clustering. Here, we observe that the assumption of a stationary correlation structure, for e.g. principal component analysis, is not justified.
\begin{table}
	\caption{Minimum, first quartile, median, mean, third quartile and maximum of silhouette coefficients for the clustering solutions with 2-5 clusters for correlation matrices of WT1.}
	\centering
	\begin{tabular}{lllllll}
		\toprule
		clusters & min & 1st Qu. & median & mean & 3rd Qu. & max \\
		\midrule
		2 & -0.046 & 0.473 & 0.571 & 0.540 & 0.664 & 0.734 \\
		3 & -0.157 & 0.379 & 0.537 & 0.491 & 0.640 & 0.718 \\
		4 & -0.262 & 0.343 & 0.508 & 0.465 & 0.631 & 0.716 \\
		5 & -0.352 & 0.314 & 0.479 & 0.439 & 0.626 & 0.707 \\
		\bottomrule
	\end{tabular}
	\label{tab:silcoefs}
\end{table}
\begin{figure}
	\centering
	\includegraphics[width=0.80\textwidth]{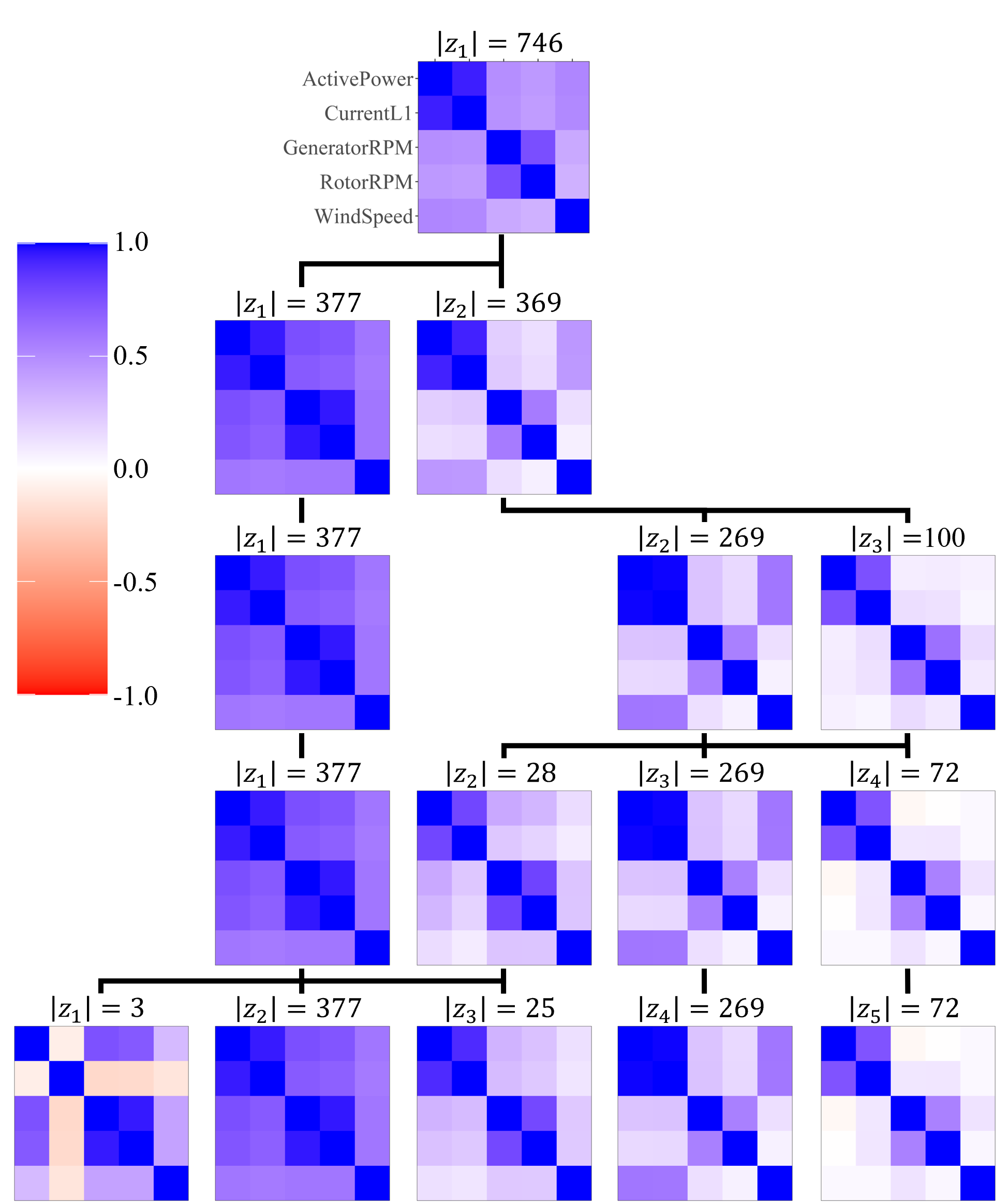}
	\caption{Cluster centroids as calculated in eq. \eqref{eq:centermatrix} for WT1 for different numbers of clusters. The color indicates the value of the correlation coefficient. Black lines connect child and parent clusters of the hierarchical algorithm and the number of cluster elements is given as $|z_n|$. Each cluster solution is ordered from low wind speeds (left) to high wind speeds (right) according to the average wind speed in a cluster.}
	\label{fig:WT1centerDendo}
\end{figure}

As mentioned before, we also look at the cluster centroids to see if the matrices show indeed different behavior and if this distinction facilitates clear identification. Figure \ref{fig:WT1centerDendo} shows the matrices calculated via eq. \eqref{eq:centermatrix} in a dendogram for the hierarchical clustering. The solutions for two and three clusters show distinctly different structures in the matrices, whereas the fourth cluster stems from cluster three, but is structurally very similar to cluster one, only differing in the strength of the mean correlation. The introduction of a fifth cluster only produces a very small cluster with only three elements. The algorithm does not identify new groups, but rather starts to classify outliers. While the average silhouette coefficient favors two clusters, we continue our analysis with three clusters as we have seen that up to this point structural differences in the matrices occur and we will later see that these can be interpreted very reasonably. Here, we also point out that structural differences in the matrix have a stronger influence on the structure of the eigenvectors, i.e. principal components, than differences in average correlation strength. They are therefore more important to distinguish when using methods like principal components and Mahalanobis distance.

\begin{figure}
	\centering
	\includegraphics[width=0.80\textwidth]{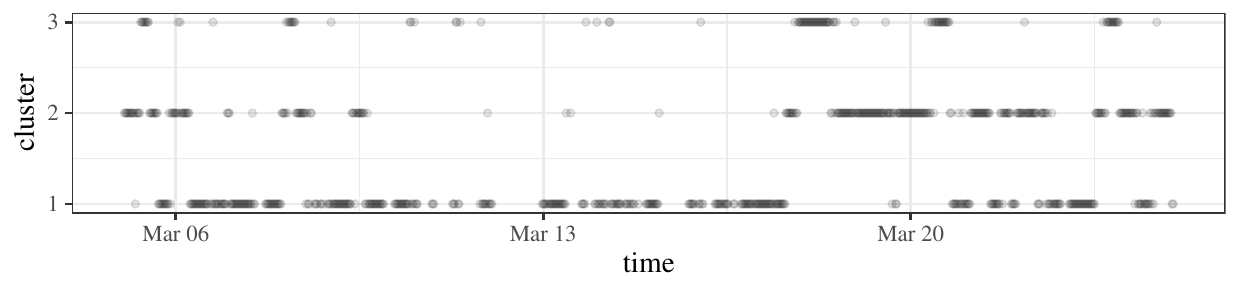}
	\caption{Cluster identifier $n$ over time for WT1. Each dot represents a 30 min epoch.}
	\label{fig:WT1clusterovertime}
\end{figure}

The classification of the matrices for three clusters is shown as an integer time series in fig. \ref{fig:WT1clusterovertime}. All three states appear to have a certain stability. Consecutive epochs often belong to the same cluster. However, there is no obvious behavior in dependency of the time. State 3 appears far less often than states 1 and 2. There is no emergence of new or disappearance of old states over time as is sometimes seen in other complex systems \cite{Munnix2012}. To get a better idea what each state might represent, we look at the matrices for the cluster centroids calculated according to eq. \eqref{eq:centermatrix} once more. They are seen in the third row of fig. \ref{fig:WT1centerDendo}. Generally, as the differences between the matrices are quite clear we can conclude - in accordance with the silhouette coefficient - that the clustering does indeed separate the matrix time series into meaningful groups. The correlation matrices are non-stationary and automatically separable with a clustering algorithm.

In every cluster the strongest correlations are clearly visible between the observable pairs ActivePower-CurrentL1 and RotorRPM-GeneratorRPM. This was to be expected as these pairs are very directly linked. Apart from this we can see that for cluster 1 both of these pairs and the WindSpeed are all correlated with each other. Put differently, higher wind speed leads to faster rotation and thus to higher power. In cluster 2 this changes and the observables RotorRPM and GeneratorRPM, while still being closely correlated with each other, decouple from the other observables. Cross-correlations between these two and any other measurement vanish. The remaining cross-correlation between WindSpeed and the two observables ActivePower and CurrentL1 also vanish in cluster 3. Clearly, the three different states of the correlation structure identified via clustering are meaningful: They show distinctly different correlation behavior between different observables.

To interpret the meaning of the clustering solution, it is helpful to look at turbine control systems. The basic functionality of such a system is, for example, described by \cite{Schutt2014}. The specific functionality varies for individual turbine types, so it is likely that not all turbine types will show the same operational states. The turbine control system of the \emph{Vestas} V90 turbines analyzed in this paper is one with variable pitch (\emph{Vestas OptiTip\texttrademark}) and variable speed (\emph{Vestas OptiSpeed\texttrademark}). We can connect the three clusters to different operational states of this turbine type, which are separated by boundary wind speeds $v_{\mathrm{on}}, v_1, v_2$ and $v_{\mathrm{nom}}$. For very low wind speeds just above the turn-on wind speed $v_{\mathrm{on}} \leq v < v_1$ the generator rotation is kept constant at the lowest possible value defined by the maximum slip in the generator. This results in a correlation structure as seen in cluster 2. Already for slightly higher wind speeds $v_1 \leq v < v_2$ the system controls the rpm proportional to the wind speed to operate at maximum aerodynamic efficiency of the rotor. This corresponds to cluster 1. With even more wind, but still not enough to reach nominal power output $v_2 \leq v < v_{\mathrm{nom}}$ the turbine operates at fixed nominal rpm by controlling the torque. The rotational variables decouple again as for very low wind speeds and the correlation structure corresponds to cluster 2 again. Of course, fluctuations in wind speed will cause the rotation to fluctuate around the nominal value leading to some noise in the correlation structure. If wind speed is high enough to allow full power output $v_{\mathrm{nom}} \leq v$ the nominal power output is reached and therefore kept constant alongside the rpm. This results in correlations as seen in cluster 3. All boundary wind speeds $v_{\mathrm{on}}, v_1, v_2, v_{\mathrm{nom}}$ are turbine dependent and usually not public knowledge.
\begin{figure}
	\centering
	\includegraphics[width=0.80\textwidth]{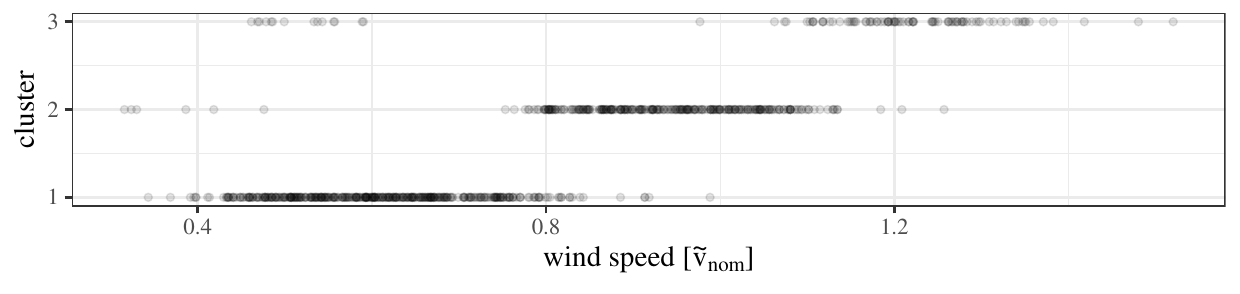}
	\caption{Cluster identifier $n$ over wind speed for WT1. Each dot represents a 30 min epoch.}
	\label{fig:WT1clusteroverwindspeed}
\end{figure}

We can see that our reasoning for the turbine at hand is correct by plotting the cluster state over the mean wind speed of the epoch instead of the time stamp in fig. \ref{fig:WT1clusteroverwindspeed}. For now disregarding the interval $v_{\mathrm{on}} \leq v < v_1$ due to lack of enough data, cluster 1 represents low wind speeds, cluster 2 intermediate wind speeds and cluster 3 high wind speeds, where nominal power output can be reached. Some exceptions are to be expected due to either wrong sorting during the clustering or wind speeds changing during the 30 min epoch. Another, but probably less important factor is the finite response time of a wind turbine controller. It ranges in seconds or minutes and therefore the correlation structure does not respond instantly to fluctuations and changes in wind speed \cite{Behnken2020}. One example are the epochs sorted into cluster 3 whose wind speeds seem to lie in the range of cluster 1. They are all moved into the fourth cluster if we take another step in the algorithm. Its centroid is shown in the dendogram fig. \ref{fig:WT1centerDendo} and exhibits a structure very close to cluster 1. Such mismatches occur due to the heuristic nature of the algorithm and noise and fluctuations in the data, which results in matrices lying on the edge between two clusters. We will see in the following section that we can use the silhouette coefficient to identify them. One might also imagine high turbulence intensities, i.e. the standard deviation of wind speed divided by the mean wind speed in an epoch, leading to strange behavior in an epoch. We have tested filtering the epochs based on this turbulence intensity and did not find significant changes in the results. The dependency on wind speed is also in accordance with the stability in time as periods with stable wind speeds are common \cite{Weber2019}. 

We have seen that the correlation matrix is non-stationary in time. The clustering has confirmed a primary influence of the control strategies in dependence of the wind speed. While the existence of different control regimes is not new, our analysis proves that they have strong influence on the structure of the correlation matrix. This automatic separation of states is a vital first step to account for non-stationarities when performing any analysis on high frequency SCADA data.

\subsection{Additional results with pitch angle and tip speed ratio}
\label{sec:WT1ClusteringAdditional}

The inclusion of the tip speed ratio does not affect the number of calculable epochs as it is directly derived from two other observables. Including the pitch angle variable, we only get 623 epochs, for which it is possible to calculate a correlation matrix. This is 123 epochs less than before. As missing values in the time series of pitch angle were filled as described in section \ref{sec:data} this can only be due to standard deviations in the pitch angle being zero for the pitch angle data. We want to point out that this can be a direct result of the filling mechanism used for missing values. This goes to show that the results with pitch angle while being interesting have to be treated with caution.

The calculation of matrices and the clustering are carried out in exactly the same way as before.

\begin{figure}
	\centering
	\includegraphics[width=0.80\textwidth]{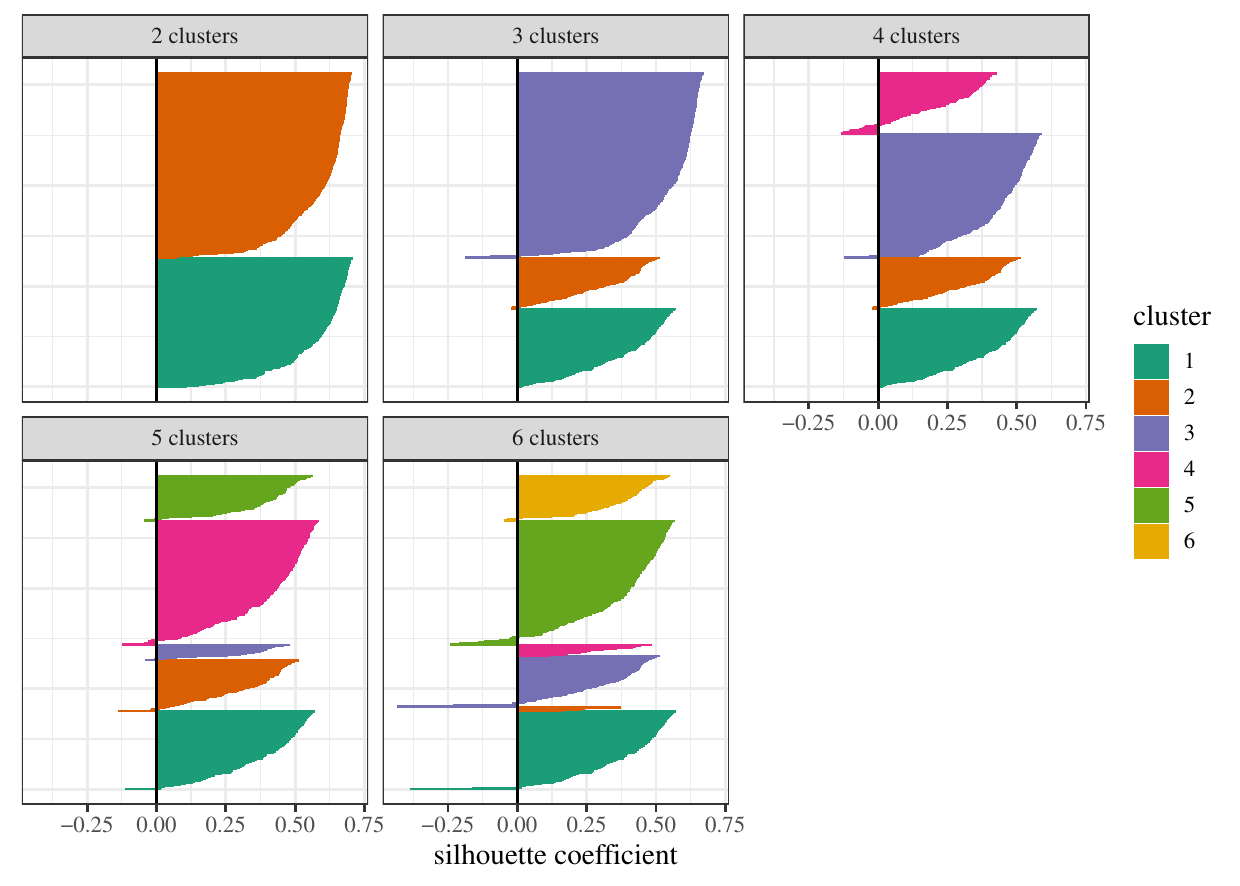}
	\caption{Silhouette plots for clustering solutions with 2-6 clusters with pitch angle. Each clustered element (matrix) is represented by a horizontal line the length of which is the silhouette coefficient for that element. Different clusters are color coded.}
	\label{fig:sil_pitch}
\end{figure}

\begin{figure}
	\centering
	\includegraphics[width=0.9\textwidth]{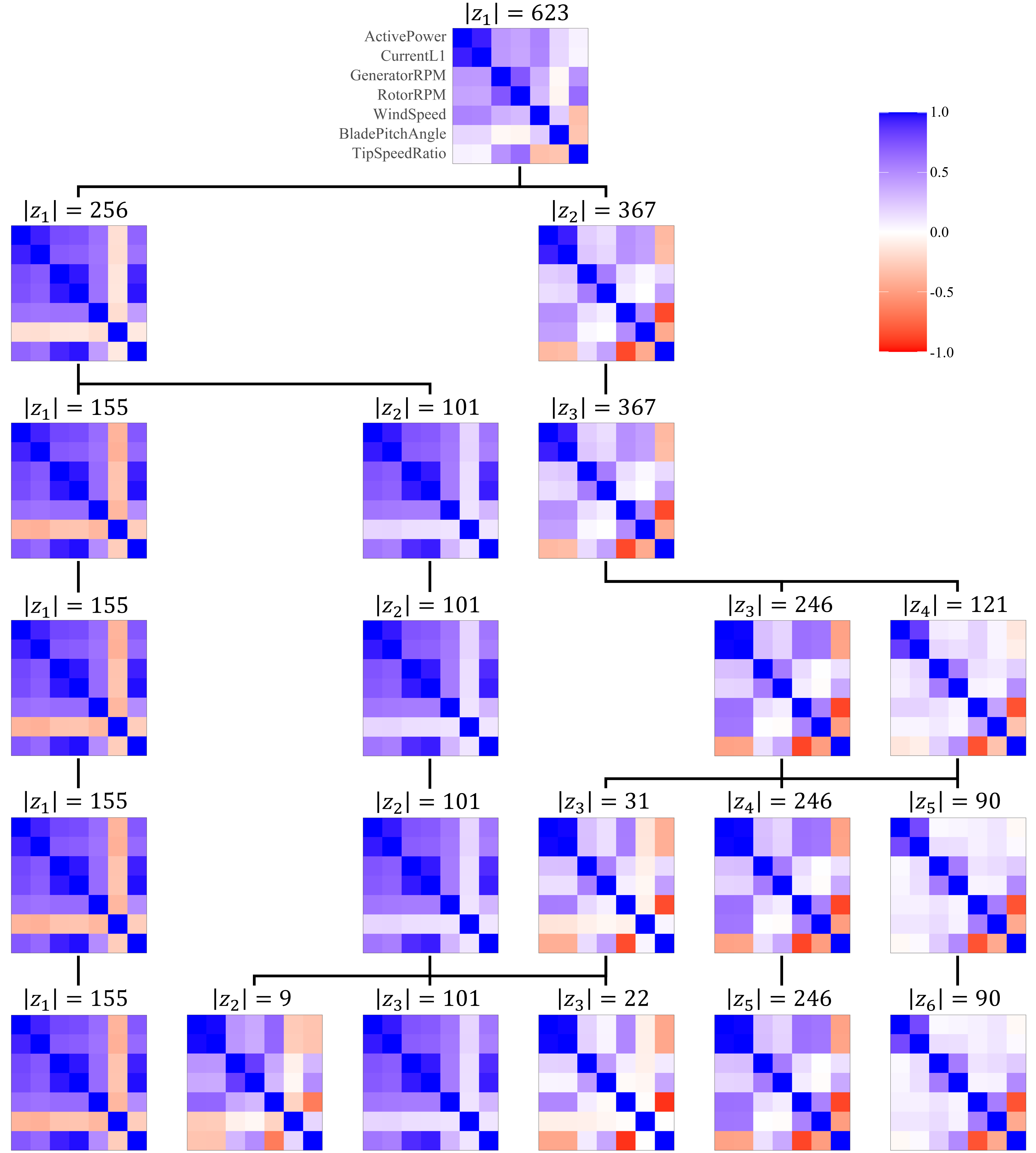}
	\caption{Cluster centroids as calculated in eq. \eqref{eq:centermatrix} for WT1 for different numbers of clusters with pitch angle. The color indicates the value of the correlation coefficient. Black lines connect child and parent clusters of the hierarchical algorithm and the number of cluster elements is given as $|z_n|$. Each cluster solution is ordered from low wind speeds (left) to high wind speeds (right) according to the average wind speed in a cluster.}
	\label{fig:WT1centerDendo_pitch}
\end{figure}

As we cannot assume that the number of relevant clusters stays the same when looking at a different set of observables, we look again at silhouettes in fig. \ref{fig:sil_pitch} and table \ref{tab:silcoefs_pitch} and the cluster dendogram in fig. \ref{fig:WT1centerDendo_pitch}.
\begin{table}
	\caption{Minimum, first quartile, median, mean, third quartile and maximum of silhouette coefficients for the clustering solutions with 2-6 clusters for correlation matrices of WT1 with pitch angle.}
	\centering
	\begin{tabular}{lllllll}
		\toprule
		clusters & min & 1st Qu. & median & mean & 3rd Qu. & max \\
		\midrule
		2 & 0.071 & 0.521 & 0.631 & 0.580 & 0.672 & 0.706 \\
		3 & -0.183 & 0.380 & 0.500 & 0.461 & 0.609 & 0.668 \\
		4 & -0.129 & 0.239 & 0.396 & 0.353 & 0.490 & 0.587 \\
		5 & -0.133 & 0.270 & 0.418 & 0.366 & 0.495 & 0.582 \\
		6 & -0.429 & 0.243 & 0.402 & 0.347 & 0.479 & 0.569 \\
		\bottomrule
	\end{tabular}
	\label{tab:silcoefs_pitch}
\end{table}

The silhouette coefficients are again largely positive with some expected negative values from imperfect clustering heuristics. On average the values of the silhouette coefficients are smaller than in the analysis with only five variables. They still indicate a good grouping. The minimum and first quartile even increase in comparison to before, which indicates less poorly sorted matrices. A slight overall decrease in silhouettes is to be expected when clustering larger matrices as more pairs of single correlation coefficients need to be compared and each of them adds fluctuations. Once again, we see decreasing silhouette coefficients for larger numbers of clusters while still indicating that the grouping is reasonable.

Looking at the centroids in fig. \ref{fig:WT1centerDendo_pitch} we see that with four clusters the three cluster solution for with five observables has reemerged. Only cluster one from the previous solution is already split again because of the pitch angle. The numbering of clusters is done based on the average wind speed in each cluster, i.e. a low cluster number indicates low wind speeds.

\begin{figure}
	\centering
	\includegraphics[width=0.80\textwidth]{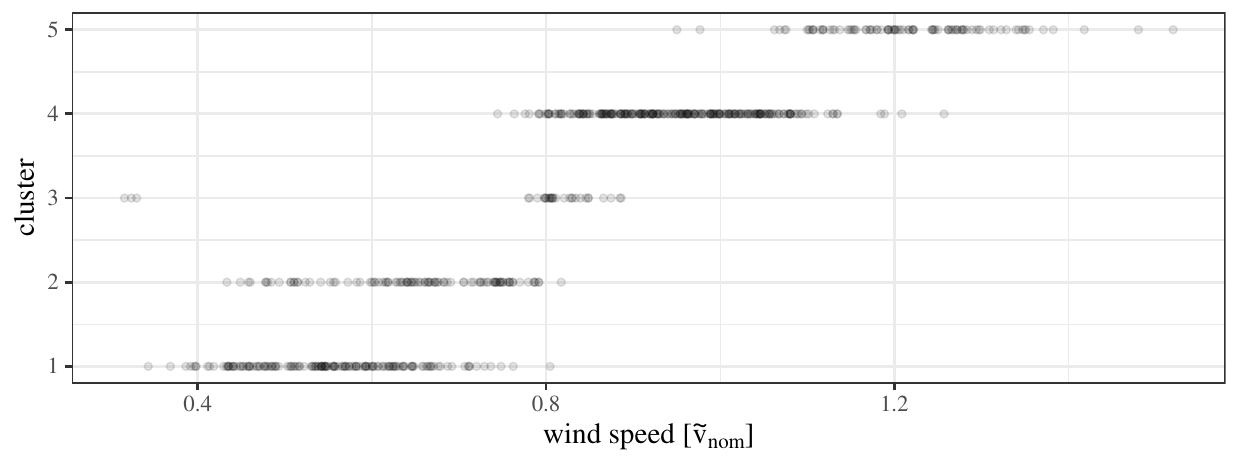}
	\caption{Cluster identifier $n$ over wind speed for WT1 and included pitch angle. Each dot represents a 30 min epoch.}
	\label{fig:WT1clusteroverwindspeed_pitch}
\end{figure}

Interestingly, with pitch angle and tip speed ratio considered we cannot stop at three clusters. The four and five cluster solution still show centroids that are structurally different. The sixths cluster distinguishes stronger and weaker values (mainly in the pitch angle) of the same type of structure and is already quite close to outsider classification with only nine inhabitants. Considering this as well as the sharply falling minimal value of silhouette coefficients from the five cluster solution to the one with six, we will take a closer look at five clusters. The cluster number over wind speed is shown in fig. \ref{fig:WT1clusteroverwindspeed_pitch}.

For low wind speeds the five main observables and the tip speed ratio are always correlated, while the pitch angle is either anti-correlated to all other observables or decouples from them. For the second case, it is likely that it simply stays constant in this regime as the intake of energy from the wind does not need to be reduced here. The anti-correlations might stem from a turbine being turned on underlined by wind speeds being slightly lower for cluster 1 than cluster 2. While it is turned off, large pitch angles are used to minimize strain on a still standing rotor and must then be reduced as wind speed increases above the cut-in point. It is not clear, however, if this effect is strong enough to produce anti-correlations for a period of 30 minutes. Cluster 3 of the five cluster solution appears to be an intermediate state, where the rotations are already decoupled from the rest of the system, but the pitch angle is not changed. Tip speed ratio is now anti-correlated to wind speed, active power and current as the rotation - and therefore the tip speed - does not increase any longer. Contrary to the average wind speed sorting cluster 3 shares a parent cluster with cluster 5 instead of 4. This might be because the decoupling of pitch angle from active power and current is a clearer distinction than the decoupling of the rotations from the rest. As the turbine controller tries to keep rotation constant, it still fluctuates creating some weak correlations whereas the pitch angle usually stays constant when it is decoupled from everything else. For intermediate wind speeds in cluster 4 the pitch angle is coupled to wind speed, active power and current. In cluster 5 at high wind speeds active power and current decouple from wind speed and pitch angle as well. In both states, the pitch angle is used to decrease the intake of power of the turbine. The tip speed ratio behaves as expected from eq. \eqref{eq:tipspeedratio}.

\begin{figure}
	\centering
	\includegraphics[width=0.8\textwidth]{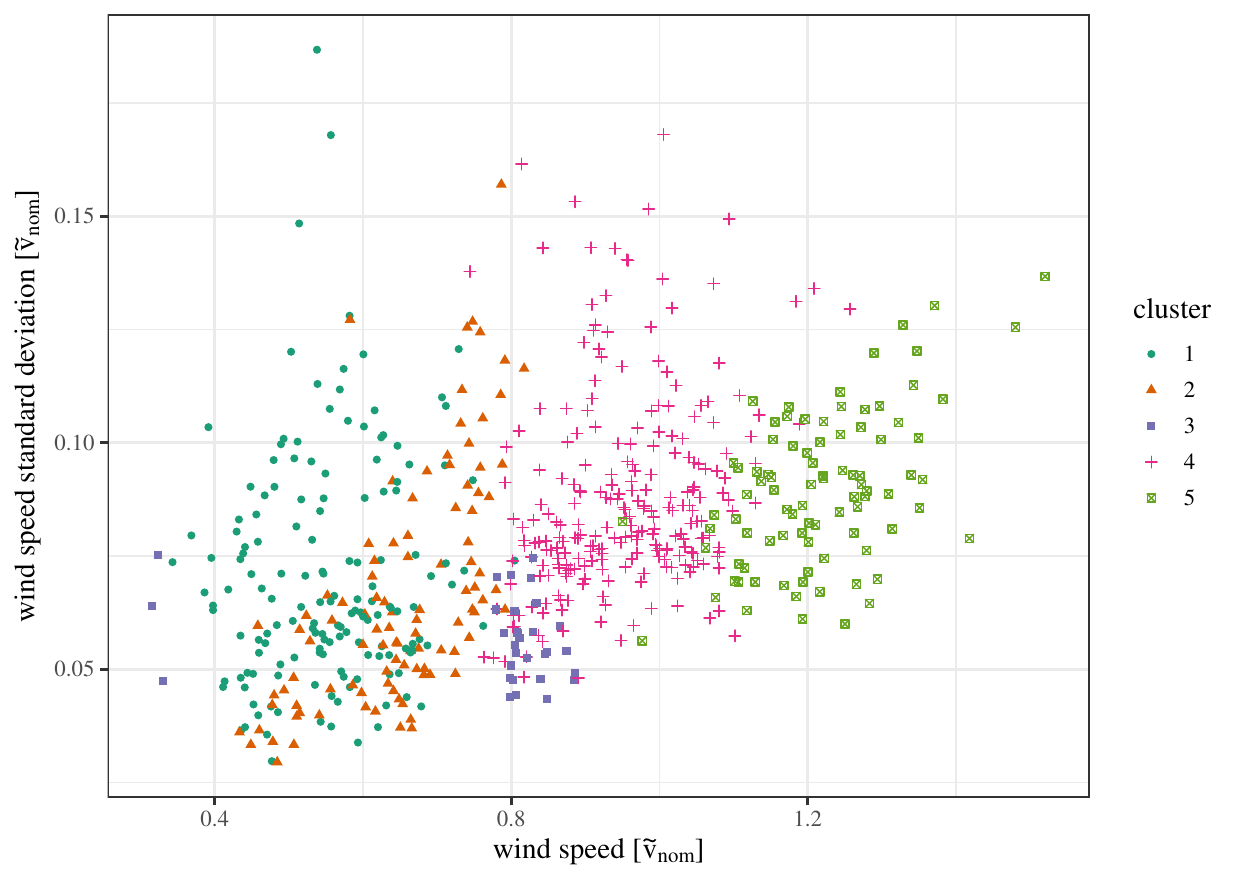}
	\caption{Cluster allocation as colored points over the average wind speed in an epoch on the x-axis and the standard deviation of wind speed during an epoch on the y-axis.}
	\label{fig:WT1clusterWindSpeedAndStddev_pitch}
\end{figure}

\begin{figure}
	\centering
	\includegraphics[width=0.8\textwidth]{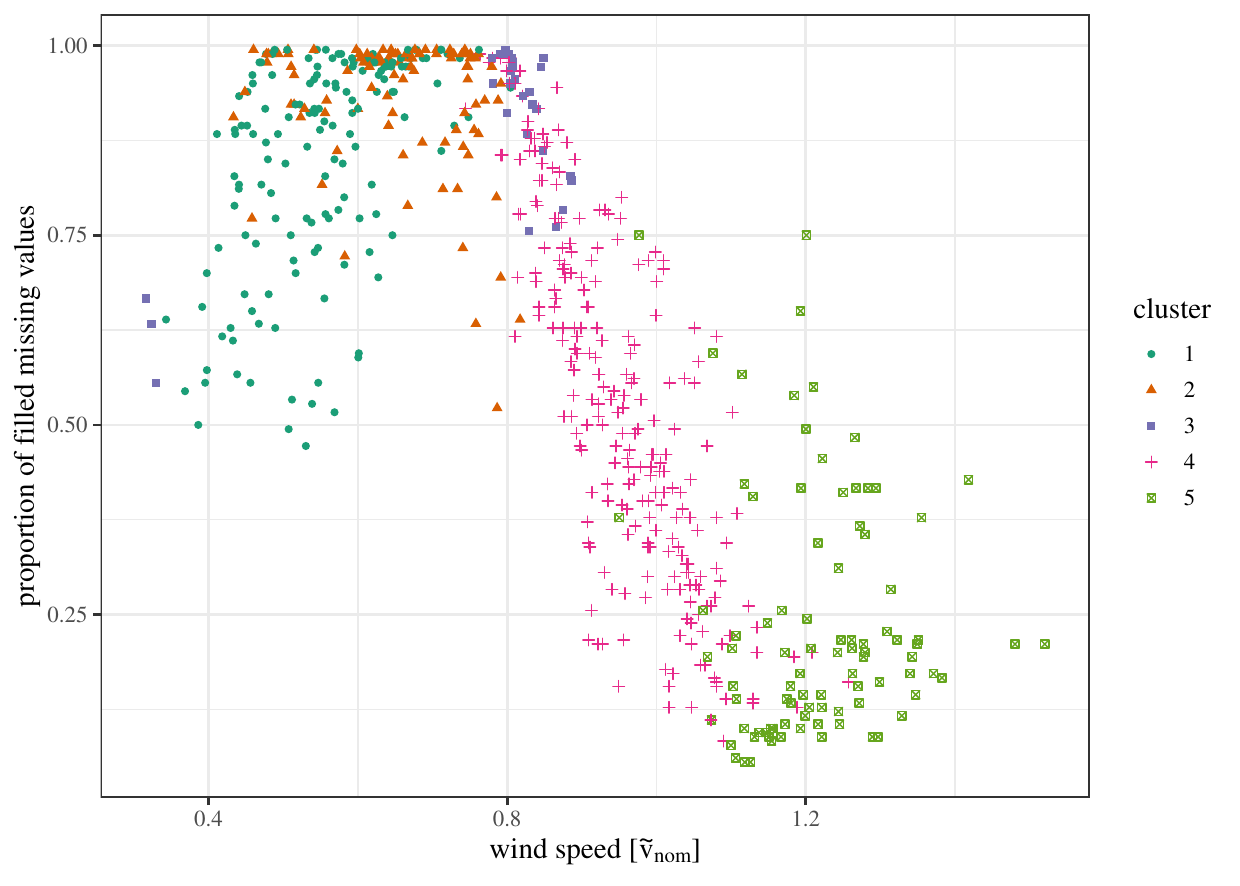}
	\caption{Cluster allocation as colored points over the average wind speed in an epoch on the x-axis. The y-axis shows the proportion of filled missing values in the pitch angle time series relative to all values in an epoch.}
	\label{fig:WT1clusterWindSpeedAndFilledPitch}
\end{figure}

Figure \ref{fig:WT1clusteroverwindspeed_pitch} shows that the clusters are not as clearly separated over wind speed alone as with only five variables. This is mainly true for low wind speeds. The three regimes we identified in our main analysis can again be distinguished. Furthermore, we see in fig. \ref{fig:WT1clusterWindSpeedAndStddev_pitch} that the small range intermediate cluster 3 can be distinguished from cluster 4 when looking not only at the average wind speed in the epoch but also at the standard deviation of wind speed in the epoch. Cluster 3 exists for small standard deviations. One possible explanation is that the wind changed so little that the controller did not change the pitch angle even though it would already do so in this regime as seen in cluster 4. There could also be a small wind speed regime where the controller already tries to keep rotation constant, but does not change the pitch angle to this end. We must also mention that the amount of filled missing values for the pitch angle time series is quite high in cluster 3 as can be seen in figure \ref{fig:WT1clusterWindSpeedAndFilledPitch}. This is also true for the overlapping clusters 1 and 2, for which we did not find a clear distinction criterion. The high amount of filled missing values in clusters 2 and 3 could be reasonable as the pitch angle is decoupled from other variables. This would lead to a constant value which would lead to many missing values in the time series, if the reasoning in section \ref{sec:data} is correct. However, without knowing for certain that this reasoning is correct, the decoupling of pitch angle from the rest that we find could also be an artifact of the data manipulation.

In conclusion, we have seen that non-stationarity can also be detected with our clustering when including pitch angle and tip speed ratio. The primary influence still stems from the control strategies in dependence of the wind. In general, it can be necessary to have more than one variable, on which the distinction between clusters is based. It could also happen that some overlap cannot be distinguished and more than one normal state would need to be considered for analysis of the system under those operating conditions.


Further analysis of matrices with more and different variables as well as an attempt to distinguish other influencing factors is interesting for future work. Regulatory impacts on the correlation structure as, for example, curtailment should also be considered. We will continue the current work with an analysis of the possibility to predict the state solely based on wind speed for our five main observables.

\section{Cluster prediction by wind speed}
\label{sec:WT1Model}

Having established a strong influence of the control system on the structure of the analyzed correlation matrices, we will now try to predict which correlation matrix state, i.e. operational state, the turbine should be in based on the wind speed. We have seen in the analysis with additional variables that overlaps between states can happen when differentiating solely based on wind speed. In such cases, more variables or external parameters might be necessary for state prediction. For the current work, we will stick with the simpler example of five observables. Here, we are confident when separating the clusters solely based on wind speed and show that such a distinction can, in general, work. This makes the example easy to follow. Also, this way we do not need to worry about the filled in pitch values. We now present a method that allows separation of the three states found based on wind speed. To this end, we look at the distribution of wind speeds in the different states, analyze them and then predict the boundary wind speeds that separate the distinct groups. This will show that it is possible to account for found non-stationarity, even though small adaptations are likely to be necessary when considering different sets of observables or turbines.
\begin{figure}
	\centering
	\includegraphics[width=0.80\textwidth]{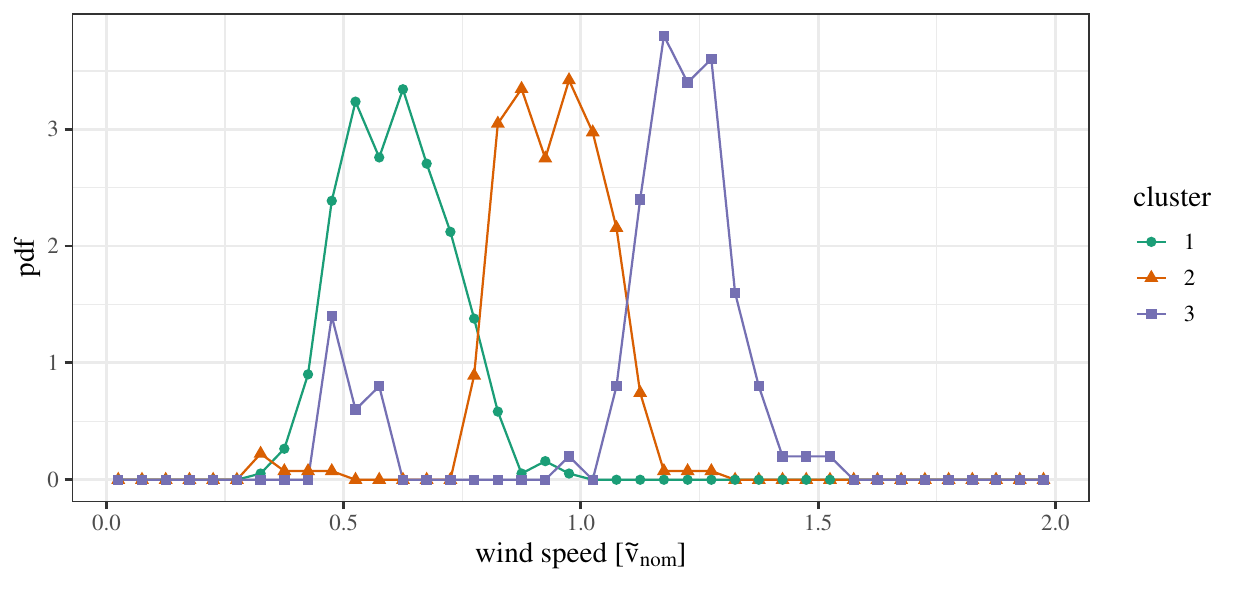}
	\caption{Probability density functions for the 30 min epoch mean wind speed per cluster. The width of bins is $0.05 \tilde{v}_{\mathrm{nom}}$.}
	\label{fig:WT1windSpeedDistr}
\end{figure}
\begin{figure}
	\centering
	\includegraphics[width=0.80\textwidth]{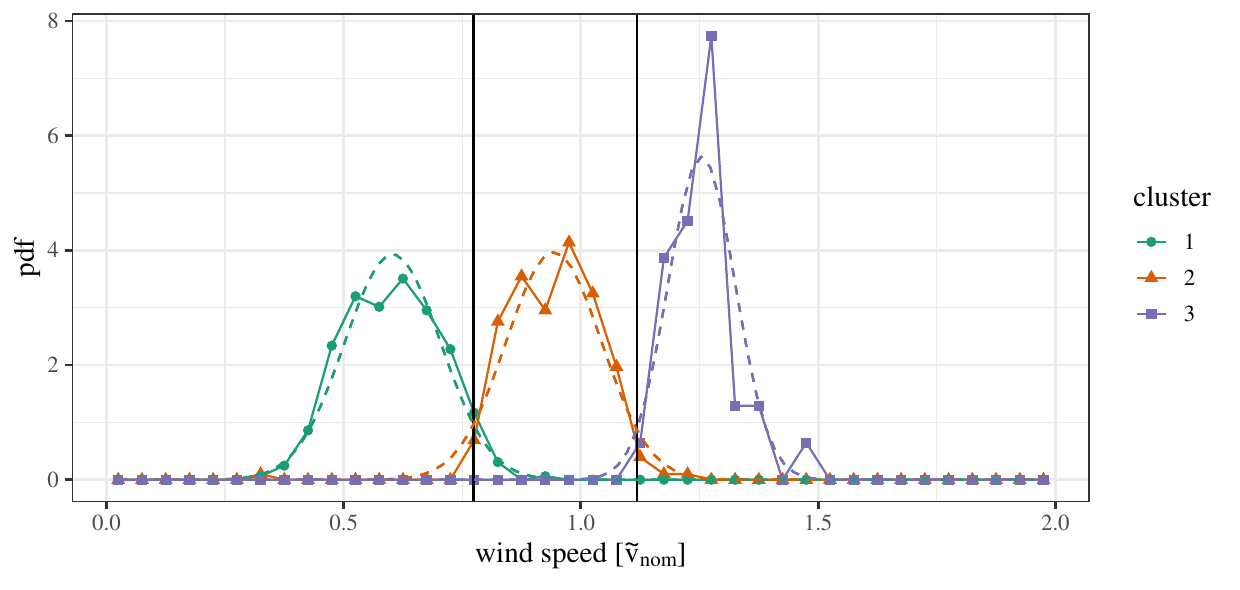}
	\caption{Probability density functions for the 30 min epoch mean wind speed per cluster using only epochs with a silhouette coefficient above the first quartile of all silhouette coefficients. For each cluster a normal distribution was fitted. Then black vertical lines indicate the intersections of these distributions and thereby the boundary wind speeds. The width of bins is $0.05 \tilde{v}_{\mathrm{nom}}$.}
	\label{fig:WT1windSpeedDistr_thr}
\end{figure}
In fig. \ref{fig:WT1windSpeedDistr} one can see the empirical probability density functions (pdf) for wind speeds per cluster state. As already expected from fig. \ref{fig:WT1clusteroverwindspeed}, we can clearly distinguish the different regimes. However, we identify much more clearly what we are calling mismatches: epochs that are sorted into cluster 3, but have mean wind speeds in the range associated with cluster 1. They make up the left peak of the distribution for cluster 3. Furthermore we can see a small peak in the probability density function for cluster 2 lying at very low wind speeds beneath the distribution for cluster 1. These could be reasonable as the rotation of the generator shaft is kept at a minimum rotational speed needed for operation of the turbine for very low wind speeds as discussed in the previous section. However, the data in the very low wind regime is sparse and not as reliable as the turbines often move in and out of operation during these times due to shutting off below a certain minimal wind speed, therefore we will disregard this first boundary $v_1$ for now.

Before modeling the boundaries between the distributions, we try to compensate for mismatches due to matrices lying at the edge of two clusters, matrices being wrongly sorted, or singular outliers. This can easily be done by using the silhouette coefficient we have introduced before. It gives an indication of how good a member of a cluster fits into this cluster compared to the other clusters. This means that any 30 min epoch that has been sorted into cluster $i$ but should rather be in cluster $j$ will have a very small or negative silhouette coefficient. We can use this fact and remove from the calculation of the probability density function all epochs with a silhouette coefficient below the first quartile of all silhouette coefficients, which can be seen in table \ref{tab:silcoefs}. The resulting probability density functions can be seen in fig. \ref{fig:WT1windSpeedDistr_thr}. The second peak at low wind speeds for cluster 3 disappears. This indicates that our reasoning of a mismatch was correct. The persistence of the small peak at very low wind speeds for cluster 2 on the other hand shows that it indeed points toward a control of the rotational speeds in this regime.

The empirical distributions are noisy due to the finite amount of data points. This is especially true for cluster 3, which contains the least epochs. However, it is very clear that every cluster is representing a wind speed regime. There are now two basic approaches to defining the boundaries between these regimes. One can simply look at the empirical data and define for each value of wind speed the maximum likelihood state based on the empirical probability density function. Secondly, one can fit a distribution to the data and calculate the intersections of these, which represent the boundaries. For now, we choose to fit distributions as it turns out that a normal distribution is a good choice for each wind speed regime (see fig. \ref{fig:WT1windSpeedDistr_thr}) and the other method could be heavily influenced by noisiness in the empirical data. Of course, this will dismiss the smaller peak of cluster 2. It should be taken into account if enough data exists in this regime (cp. section \ref{sec:multiturbine}). For now, cluster 1 has a mean of $0.603 \tilde{v}_{\mathrm{nom}}$ and a standard deviation of $0.101 \tilde{v}_{\mathrm{nom}}$. Clusters 2 and 3 are centered at $0.943 \tilde{v}_{\mathrm{nom}}$ and $1.255 \tilde{v}_{\mathrm{nom}}$ with standard deviations of $0.101 \tilde{v}_{\mathrm{nom}}$ and $0.071 \tilde{v}_{\mathrm{nom}}$ respectively. These values lead to boundaries at $v_2 = 0.774 \tilde{v}_{\mathrm{nom}}$ and $v_{\mathrm{nom}} = 1.118 \tilde{v}_{\mathrm{nom}}$. Interestingly, the last value shows that $v_{\mathrm{nom}}$ as calculated in our analysis is larger than $\tilde{v}_{\mathrm{nom}}$. The reason for this discrepancy lies in realistic operational conditions. The power curve (active power in dependency of wind speed) as given by the manufacturer is one line. Accordingly, there is exactly one value $\tilde{v}_{\mathrm{nom}}$ which marks the starting point for nominal power production. In reality, especially when looking at high-frequency data, there will always be an area around this line which is realized. The value $\tilde{v}_{\mathrm{nom}}$ lies in the middle of this smeared out power curve. At this wind speed nominal power output can be reached but is not yet constant. With even higher wind speeds, it becomes less and less likely that the actual power produced lies beneath the nominal value. Only when this probability nearly vanishes, a change in correlation structure is detectable by our method. It is therefore reasonable that our value $v_{\mathrm{nom}}$  lies higher than $\tilde{v}_{\mathrm{nom}}$. While our value is therefore well suited to distinguish correlation states, it cannot be directly compared with the nominal wind speed given by the manufacturer of a turbine. We have confirmed this by looking at scatter plots of our data but cannot show them in this paper due to data confidentiality.

We want to point out two things. First, when using our method as a pre-processing for an analysis it needs to be run on the same observables that are to be considered in the analysis (compare section \ref{sec:WT1ClusteringAdditional}). Some adaptations for additional influences from the external conditions, regulatory influences (e.g. curtailment, deration) or overlaps of clusters might be necessary. Second, it is not proven that a normal distribution will always provide the best fit. For example, if much larger wind speeds exist in the data set, the distribution for cluster 3 would have much heavier tails in the large wind speed regime unless cut off by the cut-out wind speed of the turbine.

\begin{figure}
	\centering
	\includegraphics[width=0.80\textwidth]{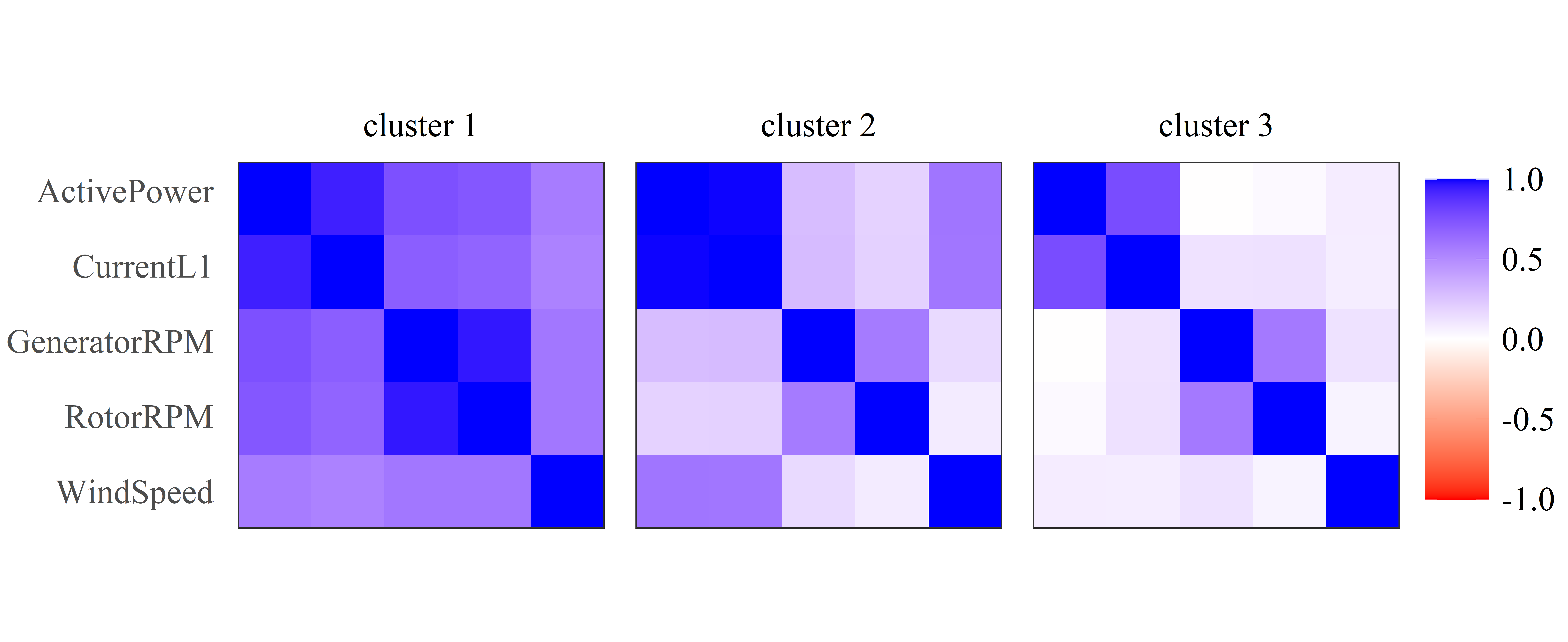}
	\caption{Matrices corresponding to the group centroids after sorting with the epochs according to the calculated boundaries for WT1. The mean matrices were calculated for all epochs, not only those used to determine the boundaries. The color indicates the value of the correlation coefficient.}
	\label{fig:WT1MatricesByModel}
\end{figure}

In the current case for WT1 the model fitted works very well. Compared to the clustering solution we have 9.9\% of changes in group assignment. If one only looks at epochs with silhouettes above the first quartile this number reduces to 3.7\%. This is clear as the epochs previously characterized as mismatches obviously change their cluster allocation when applying the model to all epochs. The mean correlation matrices of the states as sorted by the model are shown in fig. \ref{fig:WT1MatricesByModel}. They clearly exhibit the different structures discussed above produced by the control system of the turbine showing that our state prediction works.
This state prediction is an essential first step for using the different operational regimes as a pre-processing for data analysis. Using the fitted criterion, one can predict what state the turbine should be in and run the analysis for the corresponding operational regime. This is necessary if the analysis itself (e.g. principal components) directly involves the correlation matrix. Otherwise, it is also possible to make direct use of the clustering and simply identify the state by comparing the current correlation matrix to the previously identified cluster centers. A direct application of both that we want to test in the future is monitoring the correlation structure. Using the fitted criterion we predict a correlation matrix and compare it to the current one. With sufficiently labelled data we plan to analyze if deviations might signify operational problems or failures.

As we have seen in our analysis with additional variables in sec. \ref{sec:WT1ClusteringAdditional} for low wind speeds, large overlaps between multiple states can occur when differentiating by wind speed. If this is the case in an analysis at hand, one could look for other distinguishing factors. However, it is not a given that these exist. Another possibility is to accept that more than one control state is normal for the given conditions and compare new data to all possible states. If for example the goal is to minimize false alarms in a failure detection procedure, one could run failure analysis in all likely states and then choose the one that gives the least indication for failure. An alternative could be weighing the failure indicators with the likeliness calculated for each state under the given conditions.

\section{Application to multiple turbines}
\label{sec:multiturbine}

For a single wind turbine we identified different operational states in the correlation structure and presented a model to distinguish these states based on wind speed. To be useful for applications, our findings need to be general characteristics and not be specific for one turbine. We proceed and test our methodology for all turbines in the data set. We want to emphasize that in a first step this does not mean assuming one model with fixed wind speed boundaries and applying it to all turbines. Rather, we test if the procedure described in previous chapters can be automatically transferred to other turbines without supervision. Hence, we perform cluster analysis and fit the boundaries for each turbine separately.

An easily comparable indicator for the quality of the proposed methodology is the relative numbers of cluster allocation changes from the model compared to the clustering itself. We already discussed that for WT1 at the end of the previous section. This number will drastically increase if either of the two steps in the calculation does not work well on a turbine: If the clustering algorithm returns a solution that is not grouped by wind speed, sorting on the basis of wind speed will change the allocation of many epochs. If they are clustered according to wind speed, but the boundaries are less sharp than for WT1, it will again result in many changed allocations.
\begin{figure}
	\centering
	\includegraphics[width=0.48\textwidth]{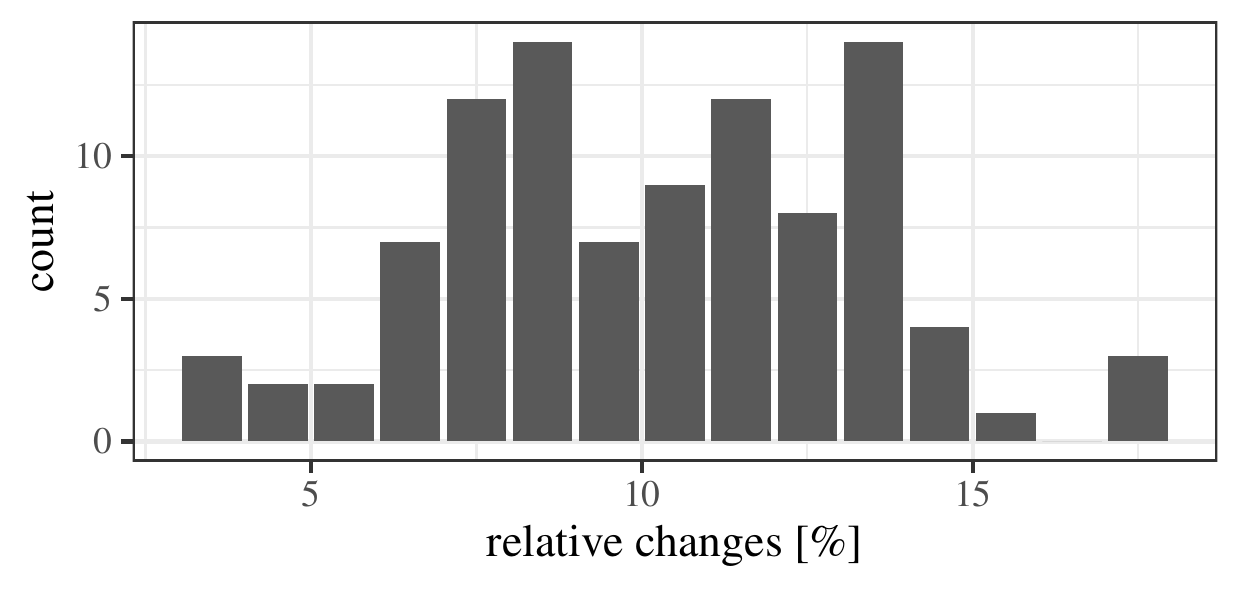}
	\caption{Histogram counts showing the frequency of relative changes in state allocation when comparing clustering and individual models per turbine.}
	\label{fig:allocationChange1}
\end{figure}
\begin{figure}
	\centering
	\includegraphics[width=0.48\textwidth]{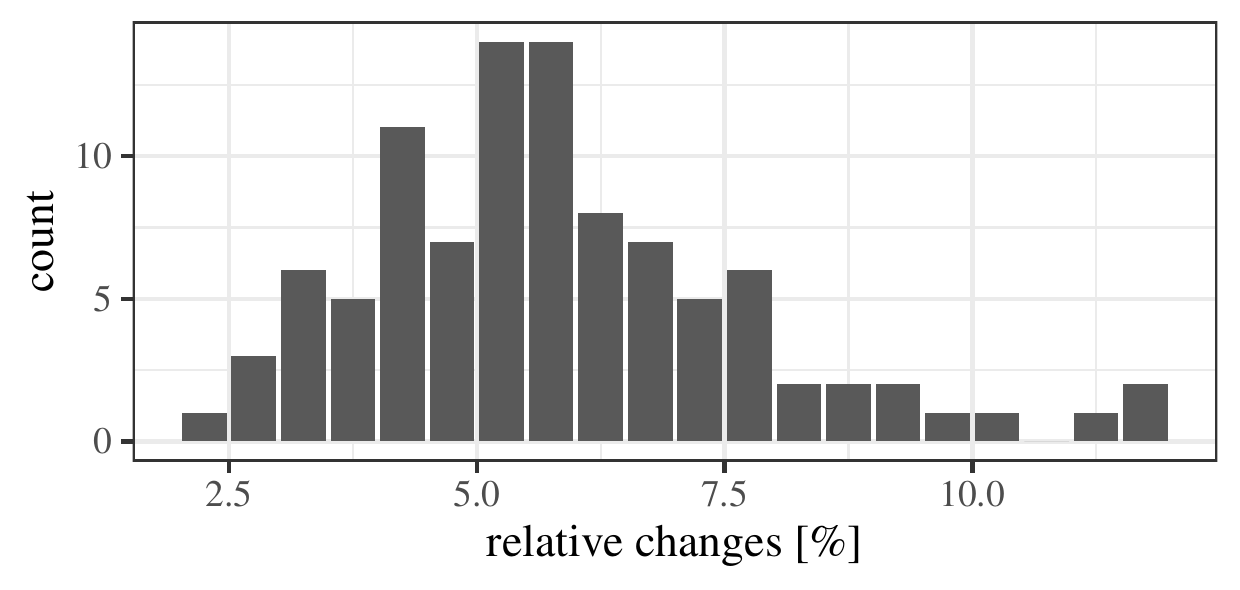}
	\caption{Histogram counts showing the frequency of relative changes in state allocation per turbine when comparing the clustering solution and the maximum likelihood model based on data from all turbines.}
	\label{fig:allocationChange2}
\end{figure}

A histogram over all calculated allocation changes for the 98 turbines is shown in fig. \ref{fig:allocationChange1}. We can see that the changes for WT1 lie in the lower end of changes, as expected due to it not showing any alarms or failures in the chosen time span. This does not hold true for the other turbines, most of which exhibit a few more changes. However, there are multiple turbines which show no more changes than WT1. Also, for those that do the fitted boundaries still work remarkably well. Some allocation changes are always expected for the reasons discussed above and especially near the boundaries the distinction between two states is not always perfect. Concluding that our proposed method works well for all turbines in our dataset, we proceed with an optimization.

\begin{figure}
	\centering
	\includegraphics[width=0.80\textwidth]{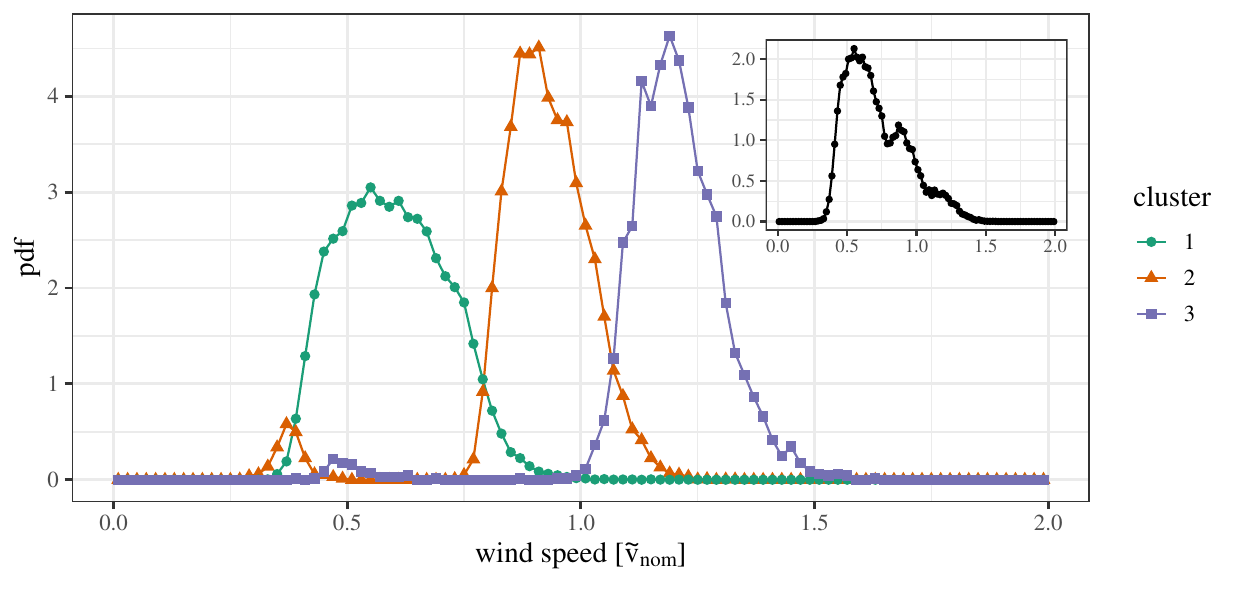}
	\caption{Probability density functions for the 30 min epoch mean wind speed per cluster considering all turbines. Only those epochs with a silhouette coefficient above the first quartile of all silhouette coefficients for each turbine were used (calculated separately per turbine). The underlying wind speed distribution without cluster separation is shown as inlay. The width of bins is $0.02\tilde{v}_{\mathrm{nom}}$.}
	\label{fig:allTurbinewindSpeedDistrCluster_thr}
\end{figure}

It stands to reason that the probability density functions for wind speeds per cluster should be much smoother, if we look at all turbines at the same time. Combining the data from all turbines before fitting the model, we have approximately 98 times more data points than for a single turbine. The resulting distributions are shown in fig. \ref{fig:allTurbinewindSpeedDistrCluster_thr}. They are indeed much smoother. Furthermore, we can see that the previously assumed Gaussian fit would not work well anymore. Especially the distribution shown for cluster 1 is skewed. Also, the peak in the distribution for cluster 2 beneath wind speeds of $0.4 \tilde{v}_{\mathrm{nom}}$ becomes more explicit. This second point is explained in sections \ref{sec:WT1Clustering} and \ref{sec:WT1Model} and actually underlines our reasoning: For very low wind speeds rotation is kept constant and with more data from all turbines we can distinguish this regime more clearly than before.

\begin{figure}
	\centering
	\includegraphics[width=0.80\textwidth]{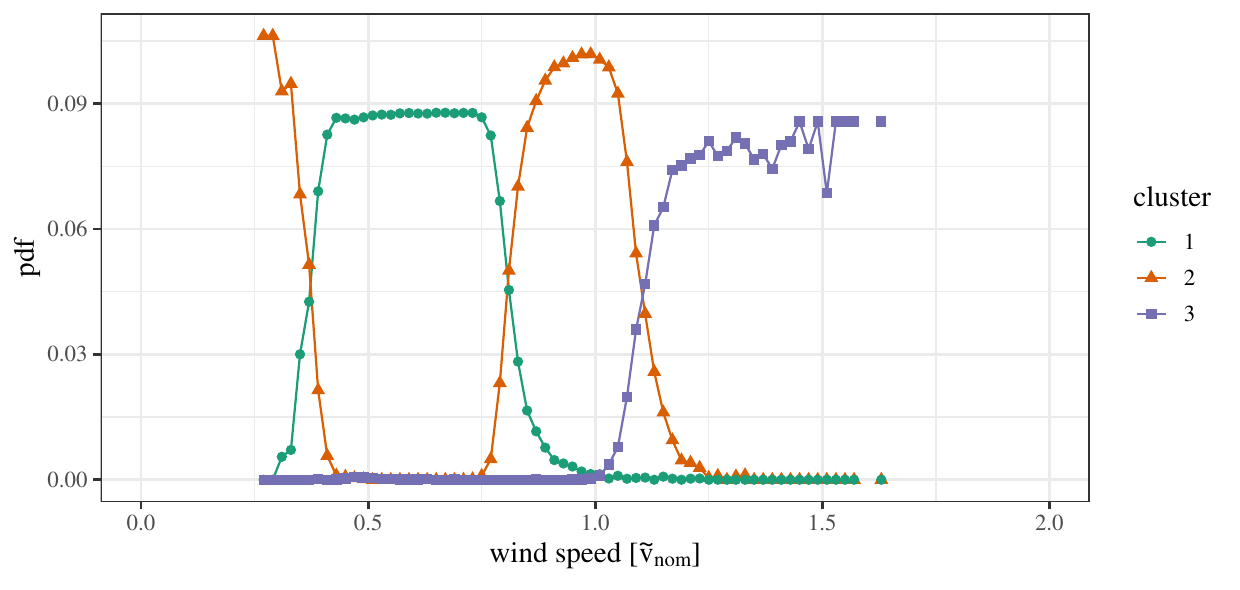}
	\caption{Probability density functions for the 30 min epoch mean wind speed per cluster considering all turbines after dividing by the total number of counts per wind speeds to rescale the underlying distribution to an equal one. Only those epochs with a silhouette coefficient above the first quartile of all silhouette coefficients for each turbine were used (calculated separately per turbine). The width of bins is $0.02\tilde{v}_{\mathrm{nom}}$.}
	\label{fig:allTurbinewindSpeedDistrCluster_scaled_thr}
\end{figure}

At a first glance, the skewness of the probability density functions does not fit our theory so well. If the controller would work perfectly and instantly and the wind speeds were constant during each epoch, we would expect the distinctions between the operational states to be represented by rectangular functions as probability densities. In non-perfect conditions these would overlap and smooth out to be something similar to a Gaussian distribution as assumed before, but they should not become skewed. However, the reason can be found in the underlying distribution of wind speeds in the inlay in fig. \ref{fig:allTurbinewindSpeedDistrCluster_thr}. It follows roughly the expected Weibull distribution. The deviations could be explained by combinations of influences in the environment of the wind farm, overlying of different Weibull distributions for different wind directions or maybe even measurement effects due to the sensor being placed behind the rotor. Some differences might also be introduced by the removal of low silhouette coefficients as these will appear often in the regimes of the boundaries between states where two correlation structures might be mixed during an epoch. This is of interest for future studies. For now, we can take away that the skewness of this underlying distributions might lead to the skewness in the cluster probability density functions. To check this we replace the histogram count of epochs $h_i(v_w)$ for wind speed $v_w$ and cluster $i$ with the rescaled count
\begin{equation*}
	\tilde{h}_i(v_w) = \frac{h_i(v_w)}{h_{\mathrm{total}}(v_w)}
\end{equation*}
by dividing with the total histogram count of epochs $h_{\mathrm{total}}(v_w)$ for that wind speed and all clusters. This basically removes the effect of the underlying wind speed distribution by transforming it into an equal distribution. The resulting probability density functions for each cluster can be seen in fig. \ref{fig:allTurbinewindSpeedDistrCluster_scaled_thr}. Indeed, we can now see symmetric areas, showing behavior very similar to rectangular functions for cluster 1 and 3 with cluster 2 being smoothed out to a more Gaussian curve, because its wind regime is quite narrowly bounded by overlaps with the other states. This strongly underlines the existence of three regimes corresponding to operational states of the turbines.

As the functions for all turbines are much smoother and the bin size can be reduced, we can apply the direct maximum likelihood method instead of fitting a continuous curve to decide cluster allocation based on the epoch wind speed. This leads to three instead of the previous two boundary wind speeds to account for the appearance of cluster 2 in the very low wind regime. The values of the boundaries are $v_1=0.38\tilde{v}_{\mathrm{nom}}$, $v_2=0.80\tilde{v}_{\mathrm{nom}}$ and $v_{\mathrm{nom}}=1.10\tilde{v}_{\mathrm{nom}}$. The resulting histogram of changes due to model allocation compared to the clustering is presented in fig. \ref{fig:allocationChange2} and shows less changes compared to fig. \ref{fig:allocationChange1}. One reason for this is the taking into account of the very low wind speed regime in cluster 2. Such a method without need for fitting could be easily transferred to other data and turbine types providing high usability as a pre-processing step for data analysis. It will need to be tested in future work, how best to deal with wind speeds where the clusters overlap. The simplest method proposed above is an all-or-nothing approach choosing the likeliest cluster. Contrary to that, if one wants to minimize false alarms in a failure detection for example, it could be useful to compare current data to all clusters which are possible for the current wind speed and choose the one indicating the least likelihood for a failure. Additional filters alongside the current wind speed can also be necessary as seen in section \ref{sec:WT1ClusteringAdditional} where the standard deviation of wind speed helped separating clusters. Also, operating measures such as curtailment might lead to temporary changes in the boundaries between clusters and thereby create the need for an additional filter. In general, when applying this method, one should always check the clustering results beforehand.

Overall we conclude that the results formerly shown for WT1 are easily transferred to multiple wind turbines. Furthermore, the model to decide state allocation based on wind speed can be optimized by taking into account more turbines and thereby more data.

\section{Conclusion}
\label{sec:conclusion}

Using a matrix distance measure and clustering algorithm formerly applied to other complex systems we were able to identify different operational states in 30 min correlation matrices of high frequency wind turbine data without prior knowledge of the control system. This demonstrates the non-stationarity of the correlation matrix for wind turbines and its automated detectability. While the states quite often exhibit stability over a certain time period, the real dependency lies with wind speed. This is expected for wind turbine control regimes. In the analysis with additional variables, the standard deviation of wind speed during a 30 min epoch was also shown to have an influence. Furthermore, it was possible to model the boundary wind speeds between the different states for the main analysis of five observables - again without knowledge about the actual parameters used in the control system. This allowed us to recreate the cluster allocation solely based on the 30 min average wind speeds. Being developed on one turbine, the method is transferred easily to multiple turbines. Results were improved by this increased amount of data. Our study shows clearly that the control system causes detectable non-stationarity in the correlation structure of high frequency wind turbine SCADA data. The automatic separation of states is important to account for this non-stationarity when analysing such data, for example to monitor a turbine during operation.

While it is of course known that the control system of the turbine changes its operational behavior based on the external influences, our analysis proves that the influence on the correlation structure of the SCADA data is significant and an automatic distinction based on the correlation matrix is possible. Therefore, assuming a stationary correlation matrix, e.g. when applying principal component analysis for dimensionality reduction on a dataset, is unjustified.

Furthermore, it could potentially be important for monitoring with high frequency SCADA-data, e.g. when applying failure detection. Especially methods directly dependent on the correlation matrices such as principal components \cite{Denis2021} and Mahalanobis distance \cite{DeMaesschalck2000} might benefit from the definitions of multiple, distinct normal states in the correlation behavior as they usually assume stationarity. They are commonly applied to wind turbines \cite{Marugan2019,Pozo2018, Yan2019, Vidal2018, Jin2021, RuizdelaHermosaGonzalez-Carrato2018}, see also the reviews mentioned in the introduction. We have recently shown that for generic correlated systems with distinct normal states, the knowledge of these states increases the sensitivity of change detection based on principal components \cite{Bette2022, Tveten2019}. This is possible via pre-processing: Using a criterion based on historical data -- wind speed in the presented case -- new or live data could be compared to the respective operational states. Charmingly, the proposed ansatz does not require changes in established techniques, it just requires their application to multiple subgroups and is therefore easily implemented.

As non-stationarity has been shown to influence failure analysis for wind turbines \cite{Zimroz2014, Bull2018}, we aim to test the benefit of our method for this use-case in future studies using SCADA-data with labeled failure events and encourage others to do so. Handling of cluster allocation at the boundaries between clusters, where overlaps in the probability distribution exist, and abnormal operating conditions such as curtailment should then be studied as well. Additionally to using the proposed pre-processing for failure analysis methods, we intend to test also a more direct application: Predicting an expected correlation matrix based on a fitted criterion and comparing it to the current one. Subsequently, deviations might point towards operational problems. Labeled data will also allow training of the model based on all intervals without abnormalities instead of a continuous time span thereby increasing the amount of available data per turbine. Furthermore, it will then be possible to study long-term non-stationarity indicated by Jin et. al.\cite{Jin2021}.

As another obvious extension of the current work, we hope to analyze the non-stationarity for other wind turbine types and different observables.

%\backmatter

\section*{Acknowledgments}
We are most grateful to \emph{Vattenfall AB} for providing the data. We acknowledge fruitful conversations with Joachim Peinke, Matthias Wächter, Christian Phillipp, Timo Lichtenstein and Anton Heckens. This study was carried out in the project \emph{Wind farm virtual Site Assistant for O\&M decision support – advanced methods for big data analysis} (WiSAbigdata) funded by the Federal Ministry of Economics Affairs and Energy, Germany (BMWi). One of us (H.M.B.) thanks for financial support in this project.

\printbibliography

\appendix

\section{Standard $\mathbf{k}$-means clustering}
\label{sec:standardkmeans}

The standard $k$-means algorithm sorts every object $O_i, ~ i = 1,\dots,N$ from a set of $N$ objects into $k$ subsets $\lbrace z_1, \dots, z_l, \dots, z_k \rbrace$. Every subset is called a cluster. Generally, the optimal number $k$ has to be determined by a separate method \cite{Jain2010, Kaufman1990}. \\
The input for the algorithm are the objects $O_i, ~ i = 1,\dots,N$ and a distance measure $d(O_i, O_j) \geq 0, ~ i,j=1,\dots,N$ as well as a method to compute the centroid of any cluster $z_l$. Note, that the distance measure must also be defined for the centroids. Then the algorithm works as follows:
\begin{enumerate}
	\item Select $k$ objects as starting cluster centroids.
	\item Assign every object to the nearest cluster based on the distance from the object to the cluster centroid.
	\item Calculate the new cluster centroids.
	\item Repeat steps 2 and 3 until no allocation changes occur.
\end{enumerate}
In this paper the objects are the correlation matrices of the subset we wish to split into two in the hierarchical approach, therefore $k=2$ always. The cluster centroids are calculated according to eq. \eqref{eq:centermatrix} and the distance is calculated by eq. \eqref{eq:distancemeasure}.

\section{Clustering results with Spearman's rank correlations}
\label{sec:spearman}

The results presented in this study are based on the Pearson correlation coefficient that has been proven useful to establish structural features in complex systems \cite{WangS2020, Gartzke2022, Munnix2012, Heckens2020a}. However, some of the variables used in our analysis, such as wind speed and active power have a well known non-linear dependency. We have therefore tested the robustness of our results, when applying the non-linear measure of Spearman's rank correlation coefficient.

To calculate the time series of Spearman correlation matrices, we rank the individual time series $S_k^{(l)}(t)$, $k=1,\dots,K$, $l=1,\dots,L$, $t=\tau,\dots,\tau+T-1$ for signal $k$ of turbine $l$ for one epoch. Then we proceed by calculating the Pearson correlation matrices for the ranked time series. The following clustering procedure and analysis is carried out in exactly the same way as for the results in sec. \ref{sec:WT1Clustering}. We present here results for the case of five variables, which can be directly compared to the results with Pearson correlation coefficients in sec. \ref{sec:WT1ClusteringMain}.

\begin{figure}
	\centering
	\includegraphics[width=0.80\textwidth]{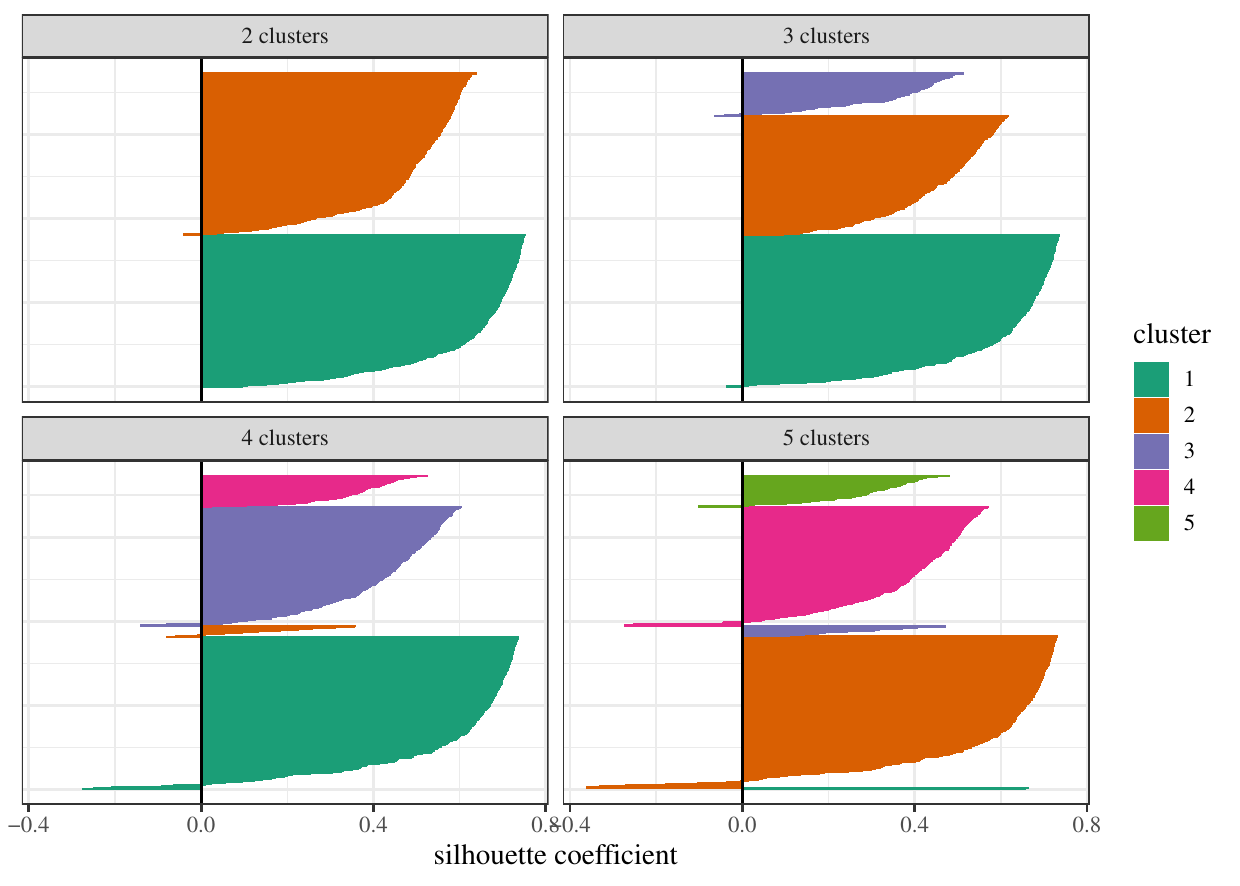}
	\caption{Silhouette plots for clustering solutions with 2-5 clusters for Spearman's rank correlation matrices. Each clustered element (matrix) is represented by a horizontal line the length of which is the silhouette coefficient for that element. Different clusters are color coded.}
	\label{fig:sil_spearman}
\end{figure}

\begin{figure}
	\centering
	\includegraphics[width=0.80\textwidth]{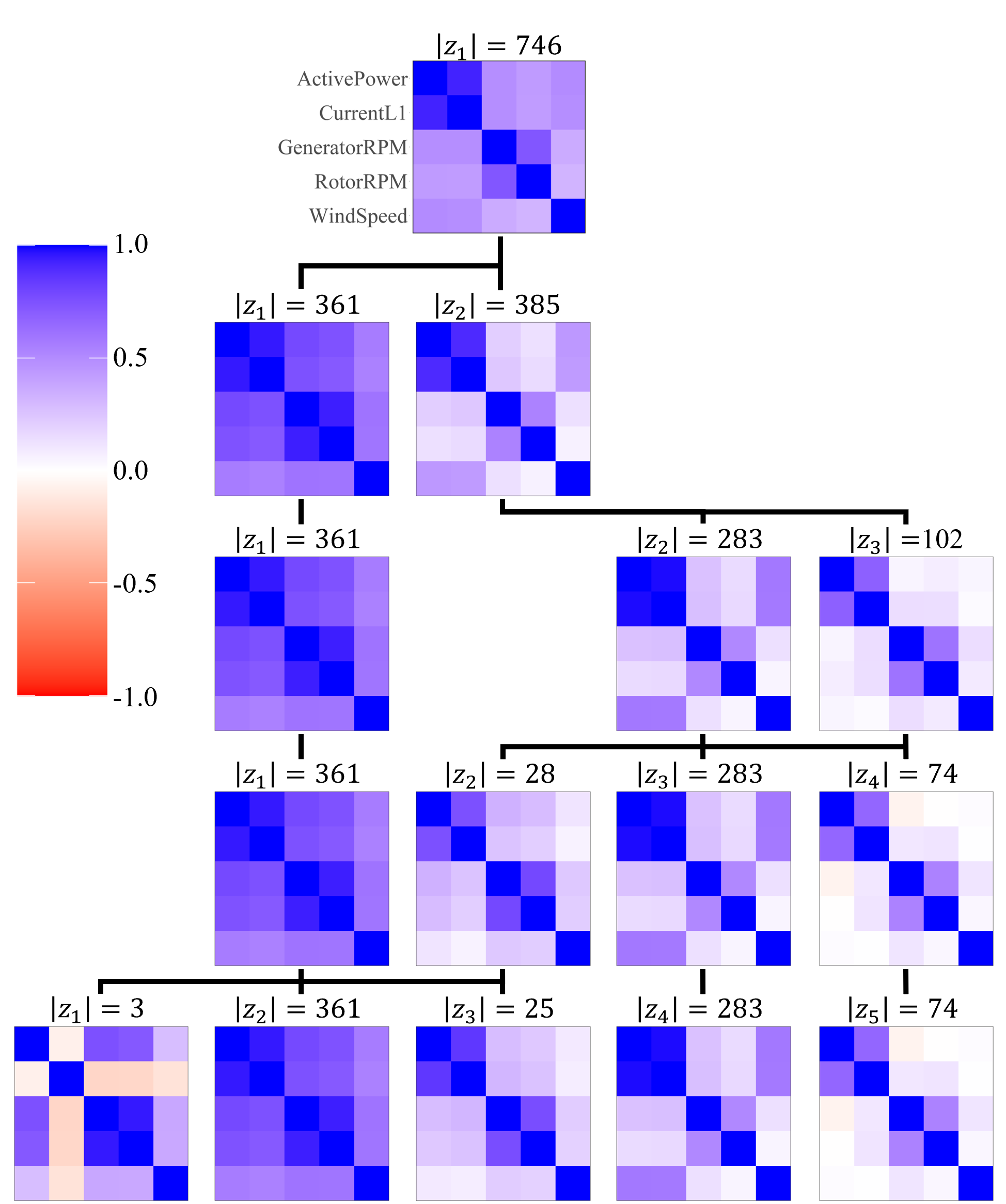}
	\caption{Spearman's rank correlation matrix cluster centroids as calculated in eq. \eqref{eq:centermatrix} for WT1 for different numbers of clusters. The color indicates the value of the correlation coefficient. Black lines connect child and parent clusters of the hierarchical algorithm and the number of cluster elements is given as $|z_n|$. Each cluster solution is ordered from low wind speeds (left) to high wind speeds (right) according to the average wind speed in a cluster.}
	\label{fig:WT1centerDendo_spearman}
\end{figure}

The silhouette coefficients shown in fig. \ref{fig:sil_spearman} and table \ref{tab:silcoefs_spearman} indicate good grouping. They show on average marginally larger values than for the Pearson correlation. Comparing the resulting cluster centers for Spearman correlations in fig. \ref{fig:WT1centerDendo_spearman} with those for simple Pearson correlations in fig. \ref{fig:WT1centerDendo}, it is obvious that the differences for the structural features revealed in this analysis are minimal. The number of elements changes slightly for some clusters, but the overall result and interpretation are the same for both correlation measures.

\begin{table}
	\caption{Minimum, first quartile, median, mean, third quartile and maximum of silhouette coefficients for the clustering solutions with 2-5 clusters for Spearman ranked correlation matrices of WT1.}
	\centering
	\begin{tabular}{lllllll}
		\toprule
		clusters & min & 1st Qu. & median & mean & 3rd Qu. & max \\
		\midrule
		2 & -0.040 & 0.479 & 0.579 & 0.550 & 0.677 & 0.751 \\
		3 & -0.064 & 0.406 & 0.533 & 0.507 & 0.656 & 0.735 \\
		4 & -0.274 & 0.365 & 0.510 & 0.476 & 0.649 & 0.735 \\
		5 & -0.361 & 0.325 & 0.483 & 0.450 & 0.641 & 0.730 \\
		\bottomrule
	\end{tabular}
	\label{tab:silcoefs_spearman}
\end{table}

\begin{figure}[h!]
	\centering
	\includegraphics[width=0.80\textwidth]{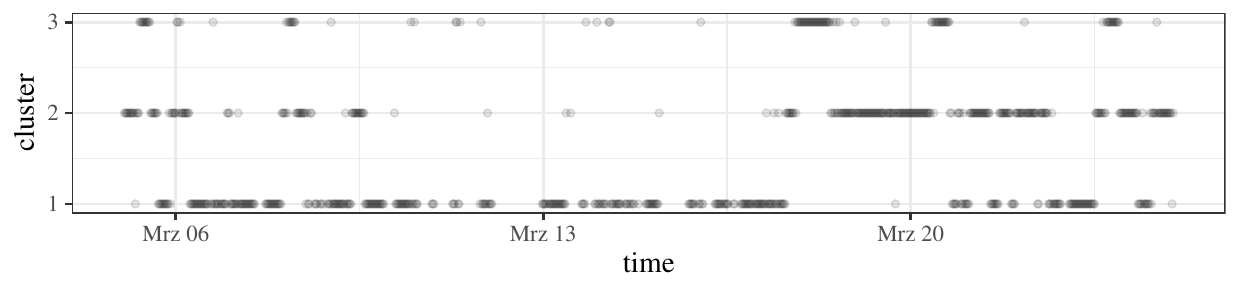}
	\caption{Cluster identifier $n$ over time for WT1 and Spearman's rank correlations. Each dot represents a 30 min epoch.}
	\label{fig:WT1clusterovertime_spearman}
\end{figure}

\begin{figure}[h!]
	\centering
	\includegraphics[width=0.80\textwidth]{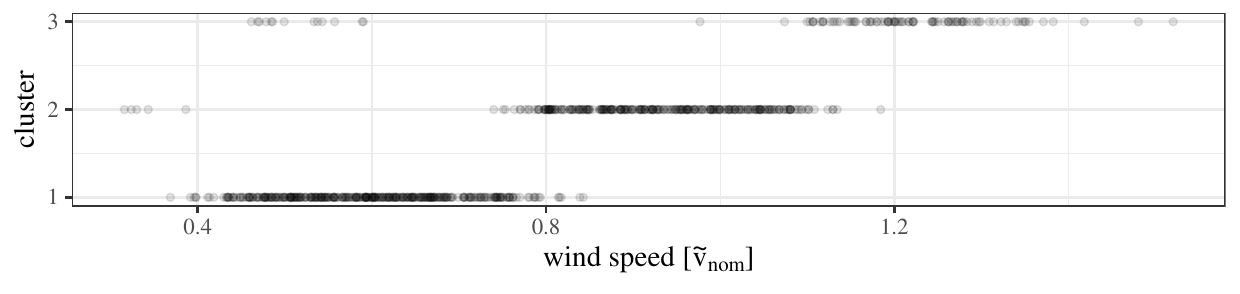}
	\caption{Cluster identifier $n$ over wind speed for WT1 and Spearman's rank correlations. Each dot represents a 30 min epoch.}
	\label{fig:WT1clusteroverwindspeed_spearman}
\end{figure}

As expected from the similarity of the results, also the plots of the cluster allocation over time in fig. \ref{fig:WT1clusterovertime_spearman} and over wind speed in fig. \ref{fig:WT1clusteroverwindspeed_spearman} are very similar to their counterparts in sec. \ref{sec:WT1ClusteringMain}.


We conclude that for the analysis carried out in this study, the simple Pearson correlation measure is sufficient. The structural differences in the correlation matrices and thereby also the structural differences in the eigenvectors, i.e. principal components, are well captured in the linear correlations.

\end{document}